\documentclass[pdflatex,sn-mathphys-ay,iicol]{sn-jnl}

\usepackage{graphicx}%
\usepackage{multirow}%
\usepackage{amsmath,amssymb,amsfonts}%
\usepackage{amsthm}%
\usepackage{mathrsfs}%
\usepackage[title]{appendix}%
\usepackage{xcolor}%
\usepackage{textcomp}%
\usepackage{manyfoot}%
\usepackage{booktabs}%
\usepackage{algorithm}%
\usepackage{algorithmicx}%
\usepackage{algpseudocode}%
\usepackage{listings}%

\newcommand{\x}{\mathbf{x}}
\newcommand{\y}{\mathbf{y}}
\newcommand{\z}{\mathbf{z}}
\newcommand{\vb}{\mathbf{v}}

\newcommand{\B}{\mathbf{B}}
\newcommand{\N}{\mathbf{N}}
\newcommand{\mub}{\boldsymbol{\mu}}
\newcommand{\gamb}{\boldsymbol{\gamma}}
\newcommand{\Psib}{\boldsymbol{\Psi}}
\newcommand{\Sigmab}{\boldsymbol{\Sigma}}
\newcommand{\var}{\mathrm{var}}
\newcommand{\cov}{\mathrm{cov}}
\DeclareMathOperator*{\argmin}{argmin}

\theoremstyle{thmstyleone}%
\theoremstyle{thmstyletwo}%

\theoremstyle{thmstylethree}%

\begin{document}

\title{Projected Normal Distribution: Moment Approximations and Generalizations}


\author*[1]{\fnm{Daniel} \sur{Herrera-Esposito}}\email{dherresp@sas.upenn.edu}

\author[1]{\fnm{Johannes} \sur{Burge}}

\affil[1]{\orgdiv{Department of Psychology}, \orgname{University of Pennsylvania}, \orgaddress{\street{Hamilton Walk}, \city{Philadelphia}, \postcode{19104}, \state{Pennsylvania}, \country{United States}}}


\abstract{
The projected normal distribution, also known as the angular Gaussian distribution, is obtained by dividing a multivariate normal random variable $\mathbf{x}$ by its norm $\sqrt{\mathbf{x}^T \mathbf{x}}$. The resulting random variable follows a distribution on the unit sphere. No closed-form formulas for the moments of the projected normal distribution are known, which can limit its use in some applications. In this work, we derive analytic approximations to the first and second moments of the projected normal distribution using Taylor expansions and using results from the theory of quadratic forms of Gaussian random variables. Then, motivated by applications in systems neuroscience, we present generalizations of the projected normal distribution that divide the variable $\mathbf{x}$ by a denominator of the form $\sqrt{\mathbf{x}^T \mathbf{B} \mathbf{x} + c}$, where $\mathbf{B}$ is a symmetric positive definite matrix and $c$ is a non-negative number. We derive moment approximations as well as the density function for these other projected distributions. We show that the moments approximations are accurate for a wide range of dimensionalities and distribution parameters. Furthermore, we show that the moments approximations can be used to fit these distributions to data through moment matching. These moment matching methods should be useful for analyzing data across a range of applications where the projected normal distribution is used, and for applying the projected normal distribution and its generalizations to model data in neuroscience.
}




\maketitle

\section{Introduction}\label{sec1}

The projected normal distribution, also known as the angular
Gaussian distribution, is a probability distribution on the
unit sphere $\mathcal{S}^{n-1}$.
A random variable $\y \in \mathcal{S}^{n-1}$ that is distributed
according to the projected normal distribution
$\y \sim \mathcal{PN}(\mub, \Sigmab)$
is obtained by radially projecting a random vector
$\x \sim \mathcal{N}(\mub, \Sigmab)$
onto $\mathcal{S}^{n-1}$ as $\y = \x/||\x||$\footnote{In the literature,
the term projected normal sometimes refers to the special
case $\Sigmab = \mathbf{I}$, where $\mathbf{I}$ is
the identity matrix. Here we use the term projected normal
distribution to refer to the distribution with arbitrary $\Sigmab$.}.
Some appealing properties of the projected normal distribution
are that it is simple to sample from, that it can be easily
generalized to arbitrary dimensions, and that
variants of the distribution can be readily obtained
by imposing constraints on the parameters of the variable $\x$
\citep{paine_elliptically_2018}.

However, the projected normal distribution has seen rare use
in the classical literature on directional statistics.
The paucity of its use is likely
due to its relatively complicated density, the fact that it has
a larger number of
parameters compared to other angular distributions
(e.g.\ the Kent distribution), and 
the lack of closed-form formulas for its moments
\citep{wang_directional_2013,paine_elliptically_2018}.
Nonetheless, interest in the projected normal distribution
is growing, due to its intuitive simplicity,
recently described variants of the distribution with
fewer parameters \citep{paine_elliptically_2018}, and
computational advances in parameterization and sampling methods
\citep{wang_directional_2013,hernandez-stumpfhauser_general_2017,yu_new_2024}.

Recent work in physics, biology, and in
Deep Learning, has used the projected normal distribution to model
data on the circle, the sphere, and the hypersphere
\citep{paine_elliptically_2018,hernandez-stumpfhauser_general_2017,
maruotti_analyzing_2016,yu_new_2024,michel_learning_2024}.
It has also recently been proposed that the projected normal distribution
is a natural choice to model the effects of divisive normalization
on response variability in neural populations \citep{herrera-esposito_analytic_2024}.
Divisive normalization is a model describing the
sublinear relation between the inputs and outputs of
a neural population \citep{carandini_normalization_2012}.
A simple version of the model
can be given by $\y = \x/||\x||$, 
where $\x$ is a vector of inputs to a neural population, and
$\y$ is the vector of the neural responses or outputs.
More flexible divisive normalization 
schemes use denominators of the form
$\sqrt{\x^T \B \x + c}$, where $\B$ is a symmetric positive
definite matrix and $c$ is a non-negative constant
\citep{carandini_normalization_2012,coen-cagli_cortical_2012,coen-cagli_impact_2013}.

\begin{figure*}[hbt]
    \centering
    \includegraphics[width=1\linewidth]{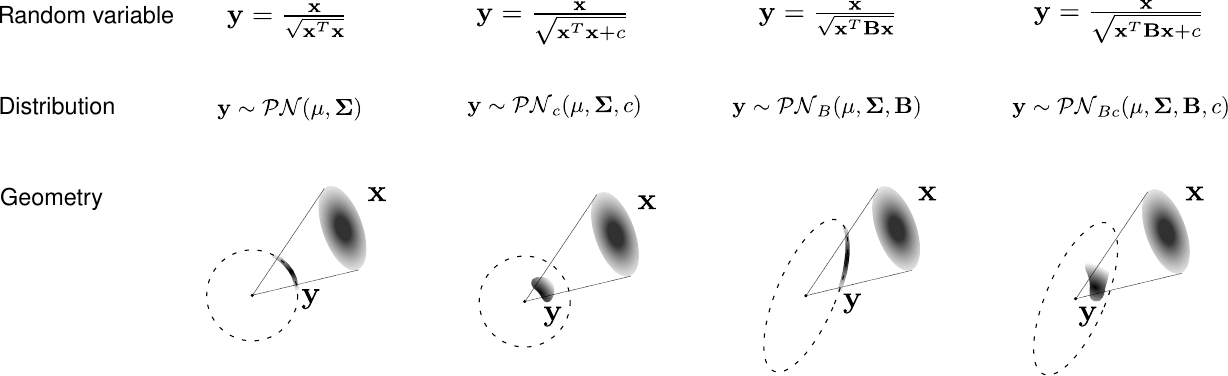}
    \caption{\textbf{Geometry of the projected normal for the 2D case,
    and the generalizations considered here.}
    Each column shows one of the four distributions
    considered in this work.
    The first row show the formula corresponding to each version
    of the corresponding random variable $\y$.
    Each distribution is obtained by dividing the Gaussian distributed
    random variable 
    $\x \sim \mathcal{N}(\mub, \Sigmab)$ by a different denominator. 
    The second row shows the notation used to refer to each version
    of the random variable $\y$. The third row shows the geometry
    of each projected variable $\y$. The first column shows the
    random variable $\y$ with a projected normal distribution,
    which projects $\x$ onto the unit sphere $\mathcal{S}^{n-1}$.
    The second column shows the random variable $\y$
    with an added constant $c$ in the denominator,
    which projects $\x$ into the interior
    of the unit sphere. The third column shows the random variable $\y$
    with a symmetric positive definite matrix $\B$ in the denominator,
    which projects $\x$ onto the surface of an ellipsoid.
    The fourth column shows the random variable $\y$
    with a symmetric positive definite matrix $\B$ and an added
    constant $c$ in the denominator, which projects $\x$ into the
    interior of an ellipsoid.}
    \label{fig:cartoon}
\end{figure*}

One particular question of interest in systems neuroscience is
how divisive normalization interacts with noise in the neural
inputs to shape the response properties of neural
population responses, especially the mean and the noise
covariance of the neural responses
\citep{goris_partitioning_2014,verhoef_attention-related_2017,
coen-cagli_relating_2019,weiss_modeling_2023,goris_response_2024}.
However, models in computational neuroscience that incorporate
divisive normalization have mostly been used to account for mean
neural responses. Response (co)variance is most often added as an
independent source of noise (but see
\cite{coen-cagli_impact_2013,goris_partitioning_2014,
coen-cagli_relating_2019,weiss_modeling_2023}).
There is a need for statistical
models that can describe the joint effects of input noise
and divisive normalization on neural population responses.
Notably, by letting the neural population inputs $\x$
be a random variable with a multivariate
normal distribution--a common noise assumption in models
of neural systems
\citep{kohn_correlations_2016,coen-cagli_relating_2019,goris_response_2024}--the
projected normal distribution and some generalized
versions of it become a natural choice for modeling how
input neural noise interacts with divisive normalization.
One limitation of the projected normal distribution
for this and other applications, however, is that no formulas
for the first and second moments of the projected normal
distribution are known.

Here, we derive analytic approximations to the first
and second moments of the projected normal
distribution $\mathcal{PN}(\mub, \Sigmab)$.
Then, motivated by applications in systems neuroscience,
we also derive moment approximations and density
functions for generalizations of the projected normal
distribution having normalizing denominators of the form
$\sqrt{\x^T \B \x + c}$,
where $\B$ is a symmetric positive
definite matrix and $c$ is a non-negative constant
(Figure~\ref{fig:cartoon}).
In all cases, we show
that the moment approximations are accurate, and
that they can be used to fit the projected normal
distribution and its generalizations to data via
moment matching. This novel fitting method and the
generalizations of the projected normal distribution
can be useful for modeling data in a range of applications,
including systems neuroscience
\citep{carandini_normalization_2012,
goris_partitioning_2014,coen-cagli_relating_2019,weiss_modeling_2023,goris_response_2024}.

\section{The projected normal distribution $\mathcal{PN}(\mub, \Sigmab)$}

\subsubsection*{Code}

All the formulas and fitting procedures presented in this
work are implemented as a Python package named
`projnormal', available at
https://github.com/dherrera1911/projnormal.


\subsubsection*{Notation}

Plain lower case letters denote scalars ($a$),
bold lower case letters denote vectors ($\mathbf{a}$), and
bold upper case letters denote matrices ($\mathbf{A}$). 
Greek letters denote moments of distributions.
An overhead bar $\bar{\cdot}$ is used to denote
the mean of a random variable, $\var(\cdot)$
to denote the variance, and $\cov(\cdot,\cdot)$
to denote the covariance between two random variables.
A distribution parameter with a hat
(e.g.\ $\widehat{\Sigmab}$) denotes a model parameter
fitted to data, rather than the true underlying parameter
of the distribution. A moment of the distribution with
a tilde (e.g.\ $\widetilde{\gamb}$) denotes a moment
approximated using the formulas derived in this work.


\subsection{Probability density function of $\mathcal{PN}(\mub, \Sigmab)$}

The density function of $\y = \x/\|\x\| \in \mathcal{S}^{n-1}$, where
$\x\sim\mathcal{N}(\mub, \Sigmab)$, is given by 
\citep{pukkila_pattern_1988,paine_elliptically_2018}:
\begin{equation}
\label{eq:pdf1}
  p(\y;\mub, \Sigmab) =
  \frac{
    e^{{\frac{1}{2}
    \left(q_2^2/q_3 - q_1 \right)}}
  }{
  \det(2\pi \Sigmab)^{\frac{1}{2}}
  q_{3}^{\frac{n}{2}}
  }
\mathcal{M}_{n-1}\left( \frac{q_2}{q_3^2} \right)
\end{equation}
where $q_1 = \mub^T \Sigmab^{-1} \mub$, 
$q_2 = \mub^T \Sigmab^{-1} \y$, $q_3 = \y^T \Sigmab^{-1} \y$,
and $\mathcal{M}_{n-1}(\alpha)$ is a function
computed recursively as
\begin{equation*}
\mathcal{M}_{n+1}(\alpha) = \alpha \mathcal{M}_{n}(\alpha) +
    n\mathcal{M}_{n-1}(\alpha)
\end{equation*}
with the initial values $\mathcal{M}_{0} = \Phi(\alpha)$ and
$\mathcal{M}_{1} = \alpha \Phi(\alpha) + \phi(\alpha)$.
$\phi(\cdot)$ and $\Phi(\cdot)$ are the standard normal
probability density function and cumulative probability function,
respectively. This formula has been used to fit the projected normal
parameters $\mub$ and $\Sigmab$ to data with maximum likelihood
through iterative optimization \citep{paine_elliptically_2018}.


\subsection{Approximation to the moments of $\mathcal{PN}(\mub, \Sigmab)$}
\label{sec:moments}

To approximate the first and second moments of
$\y \sim \mathcal{PN}(\mub, \Sigmab)$ we use Taylor expansions,
and formulas for the moments of quadratic forms
in Gaussian random vectors.
We denote the first moment and the covariance of $\y$
as $\gamb=\mathbb{E}\left[\y\right]$ and
$\Psib = \mathbb{E}\left[(\y-\gamb) (\y-\gamb)^T\right]$,
respectively.

\subsubsection{First moment approximation}
\label{sec:moments_first}

To approximate the element $\gamb_i = \mathbb{E}\left[\y_i\right]$
of the first moment, we define the auxiliary variable
$\z_i=\x^T \x - \x_i^2$, and consider the
function $\y_i = f(\x_i,\z_i)=\x_i/\sqrt{\x_i^2 + \z_i}$.
The function $f(\x_i,\z_i)$ has relatively simple
second derivatives, and the variable $\z_i$ has known moments,
allowing us to use Taylor expansions to approximate
the first moment $\gamb_i = \mathbb{E}\left[f(\x_i,\z_i)\right]$.
Taking the expectation of the
second-order Taylor expansion of $f(\x_i,\z_i)$
around the point $[\mub_i, \bar{\z}_i]$ gives
\begin{align}
\label{eq:taylor1}
    \gamb_i &= \mathbb{E}\left[ f(\x_i,\z_i) \right] \approx
        \frac{\mub_i}{\sqrt{\mub_i + \bar{\z}_i}} \nonumber \\
        &+ \frac{\Sigmab_{ii}}{2}\cdot \frac{\partial^2 f}{\partial \x_i^2}
          \Bigr|_{\substack{\x_i=\mub_i\\\z_i=\bar{\z}_i}}
        + \frac{\var(\z_i)}{2} \cdot \frac{\partial^2 f}{\partial \z_i^2}
          \Bigr|_{\substack{\x_i=\mub_i\\\z_i=\bar{\z}_i}} \nonumber \\
        &+ \cov(\x_i,\z_i) \cdot \frac{\partial^2 f}{\partial \x_i \partial \z_i} 
          \Bigr|_{\substack{\x_i=\mub_i\\\z_i=\bar{\z}_i}}
\end{align}
Solving for the second derivatives yields
\begin{align}
\label{eq:taylor2}
  \gamb_i &\approx 
    \frac{\mub_i}{\sqrt{\mub_i^2 + \bar{\z}_i}} +
    \frac{\Sigmab_{ii}}{2} \cdot \left( \frac{-3  \mub_i \bar{\z}_i}{
      ( \mub_i^2 + \bar{\z}_i)^{5/2}} \right)
    \nonumber \\
    &\quad + \frac{\var(\z_i)}{2} \cdot
    \left( \frac{3 \mub_i}{4( \mub_i^2 + \bar{\z}_i)^{5/2}} \right) \nonumber \\
    & \quad + \cov(\x_i,\z_i) \cdot \left( \frac{ \mub_i^2 -
    \frac{1}{2}\bar{\z_i}}{( \mub_i^2 + \bar{\z}_i)^{5/2}} \right)
\end{align}

The variable $\z_i$ is a quadratic form of a Gaussian random
variable, and there are known formulas for
$\bar{\z_i}$, $\var(\z_i)$ and $\cov(\x_i,\z_i)$
\citep{mathai_quadratic_1992} (see Appendix~\ref{sec:quadratics}).
Denoting the vector $\mub$ without the
$i$-th element as $\mub_{-i}$, and
the covariance matrix $\Sigmab$ without the $i$-th row and column
as $\Sigmab_{-i}$, we have
\begin{align*}
  \bar{\z}_i &= \mathrm{tr}(\Sigmab_{-i}) + \mub_{-i}^{T}  \mub_{-i}
     \nonumber \\
  \var(\z_i) &=  2 \mathrm{tr}( \Sigmab_{-i} \Sigmab_{-i})  +
  4 \mub_{-i}^T \Sigmab_{-i} \mub_{-i} \nonumber \\
  \cov(\x_i, \z_i) & = 2 \left(\mub^T \mathrm{col}_i(\Sigmab) - \mub_i \Sigmab_{ii} \right)
\end{align*}
where $\mathrm{col}_i(\Sigmab)$ is the $i$-th column of $\Sigmab$.

With straightforward algebraic manipulations,
a vectorized formula for the first moment can
be obtained (see Appendix~\ref{sec:vectorized}).

\subsubsection{Second moment and covariance approximation}
\label{sec:moments_second}

Next, we approximate the (non-centered) second moment matrix.
To approximate the $i,j$ element
$\mathbb{E}\left[\y_i \y_j\right] = \mathbb{E}\left[\x_i\x_j/\x^T\x\right]$,
consider the two variables $n_{ij} = \x_i \x_j$ and $d = \x^T \x$,
and the function
$\y_i \y_j = g(n_{ij},d) = n_{ij}/d$.
The variables $n_{ij}$ and $d$ are both quadratic forms of
the Gaussian random variable $\x$, and
$g(n_{ij},d)$ is a ratio of quadratic forms.
Computing the expected value of ratios of quadratic
forms is a well-studied problem, and the following second-order
Taylor approximation around the point $[\bar{n}_{ij}, \bar{d}]$
is reported to have high accuracy \citep{paolella_computing_2003}.
\begin{align}
  \label{eq:taylor2nd}
  \mathbb{E}[\y_i \y_j] & = \mathbb{E}[g(n_{ij},d)] \approx \nonumber \\
  & \frac{\bar{n}_{ij}}{\bar{d}} \left( 1 - \frac{\cov(n_{ij},d)}{\bar{n}_{ij}\bar{d}} +
\frac{\var(d)}{\bar{d}^2} \right)
\end{align}

The moments for $n_{ij}$ and $d$ can be computed using
formulas for the moments of quadratic forms
\citep{mathai_quadratic_1992}, resulting in
\begin{align*}
  \bar{n}_{ij} &= \Sigmab_{ij} + \mub_i \mub_j \nonumber \\
  \bar{d} &= \mathrm{tr}(\Sigmab) + \mub^T \mub \nonumber \\
  \var(d) &= 2 \, \mathrm{tr}(\Sigmab \Sigmab) + 4 \mub^T \Sigmab \mub \nonumber \\
  \cov(n_{ij},d) &= 2 \mathrm{tr}(\Sigmab \mathbf{A}^{ij} \Sigmab) +
  4 \mub^T \mathbf{A}^{ij} \Sigmab \mub
\end{align*}
where $\mathbf{A}^{ij}$ is an $n$-by-$n$ matrix where all
elements are $0$, except the
elements $ij$ and $ji$ that are equal to $1/2$ in the case
$i \neq j$, and $1$ in the case $i = j$. 
Vectorized formulas can be obtained for computing
the full second moment matrix (see Appendix~\ref{sec:vectorized}).

Finally, the covariance matrix $\widetilde{\Psib}$
is obtained using the formula
$\Psib = \mathbb{E}[\y\y^T] - \gamb \gamb^T$,
with the approximations for the first and second moments above.

\subsubsection{Exact moments for $\mathcal{PN}(\mub, \sigma^2\mathbf{I})$}

The special case where $\Sigmab = \sigma^2 \mathbf{I}$ is
important in the literature \citep{pewsey_recent_2021}.
For this special case, we derived the following
exact formulas for the mean and covariance of $y$
\begin{equation}
    \gamb = a \cdot \mub
\end{equation}
\begin{equation}
    \Psib = b \cdot{\mub \mub^T} + c \cdot \mathbf{I}
\end{equation}
where
\begin{align*}
  a &= \frac{\Gamma\left(\frac{n}{2}+\frac{1}{2}\right)}
{\sqrt{2\sigma^2} \Gamma\left(\frac{n}{2}+1\right)}
{}_1\mathrm{F}_1\left(\frac{1}{2}; \frac{n+2}{2}; -\frac{\|\mub\|^2}{2\sigma^2}\right) \\
  b &= \frac{1}{\sigma^2(n+2)} {}_1\mathrm{F}_1\left(1; \frac{n+4}{2}; -\frac{\|\mub\|^2}{2\sigma^2}\right) - a^2 \\
  c &= \frac{1}{n} {}_1\mathrm{F}_1\left(1; \frac{n+2}{2}; -\frac{\|\mub\|^2}{2\sigma^2}\right)
\end{align*}
and ${}_1\mathrm{F}_1\left(a; b; z\right)$ is the confluent hypergeometric
function. See derivation in the Appendix~\ref{sec:exact_moments}.

\section{Accuracy of moments approximation for $\mathcal{PN}(\mub,\Sigmab)$}

\subsection{Approximation accuracy for $\mathcal{PN}(\mub,\Sigmab)$}

In this section, we test the accuracy of the
approximated moments, denoted $\widetilde{\gamb}$ and
$\widetilde{\Psib}$, against
the true moments $\gamb$ and $\Psib$ of the distribution.

\subsubsection{Methods}
\label{sec:moment_methods}

For a given projected normal distribution
$\mathcal{PN}(\mub, \Sigmab)$, we obtained the
true moments $\gamb$ and $\Psib$ by sampling $10^6$
points from the distribution and computing their
mean and covariance.
To evaluate the approximations obtained with the
formulas, we compared their results to the true moments
obtained by sampling.
We evaluated the approximations for distributions with
dimensionality $n=3$, $n=6$, $n=12$, $n=24$, and $n=48$,
and different values of a variance scale parameter $s$ that
controls the overall variance (see below for details).

We evaluated the approximation accuracy in two ways.
The first is the relative mean squared error across
the elements of the distribution moments. This metric
is obtained by computing the sum of the squared
differences between the approximated and true moments,
and dividing it by the sum of the squared elements
of the true moments:
\begin{equation*}
\mathrm{Error}_{\gamb} \% = 100 \cdot \frac{\|\widetilde{\gamb}- \gamb\|^2}{\|\gamb\|^2}
\end{equation*}
\begin{equation*}
\mathrm{Error}_{\Psib}\% = 100 \cdot \frac{\|\widetilde{\Psib}- \Psib\|_F^2}{\|\Psib\|_F^2}
\end{equation*}
where $||\cdot||_F$ is the Frobenius norm.

The second metric is the cosine similarity
between the approximated and the true moments
\begin{equation*}
  \mathrm{cosine}_{\gamb} = \frac{\widetilde{\gamb} \cdot \gamb}
  {\|\widetilde{\gamb}\| \|\gamb\|}
\end{equation*}
\begin{equation*}
  \mathrm{cosine}_{\Psib} = \frac{\mathrm{tr}\left(\Psib \widetilde{\Psib}\right)}
  {\|\widetilde{\Psib}\|_F \|\Psib\|_F}
\end{equation*}

The cosine similarity is a measure of the angle between two
vectors. It is equal to $1.0$ when the vectors are
parallel, and $0.0$ when they are orthogonal. A cosine
similarity close to $1.0$ indicates that the structure of the
approximated and the true moments
is similar, without taking into account their scale.

For each dimensionality $n$ and variance scale $s$, we
sampled $100$ pairs of parameters $\mub$ and $\Sigmab$.
The parameter $\mub$ was uniformly sampled on the unit sphere.
The parameter $\Sigmab$ was obtained
by randomly sampling a diagonal matrix $\mathbf{D}$ with
positive entries (the eigenvalues of $\Sigmab$) and a random
orthonormal matrix $\mathbf{V}$ (with the eigenvectors as columns),
such that $\Sigmab = \mathbf{V} \mathbf{D} \mathbf{V}^T$.

The elements of $\mathbf{D}$ were first sampled independently from
the exponential distribution\footnote{
An alternative sampling method where the eigenvalues were 
sampled from the uniform distribution $\mathcal{U}(0.05 , 1.0)$
instead of the exponential distribution was also tested,
leading to similar results.} $\mathrm{Exp}(1)$. Then
a constant of $0.01$ was added for stability, and $\mathbf{D}$ was
scaled by a factor $s^2/n$. The scalar $s$ controls
the overall scale of the covariance matrix $\Sigmab$.
Division by $n$ was used to keep a constant ratio between the
average value of $\mathbf{D}_{ii}$ and the average squared
value $\mub^2$, the latter of which scales
with $1/n$ because of the constraint $\|\mub\|=1$.
We used values of $s=0.125$, $s=0.25$, $s=0.5$ and $s=1.0$.

To sample the eigenvectors in the columns of $\mathbf{V}$,
we first generated a skew symmetric
matrix $\mathbf{S}$, where each lower diagonal element was
independently sampled from a normal distribution. Then
$\mathbf{V}$ was obtained by taking the matrix exponential
of $\mathbf{S}$.

\begin{figure}[tbp]
    \centering
    \includegraphics[width=1.0\linewidth]{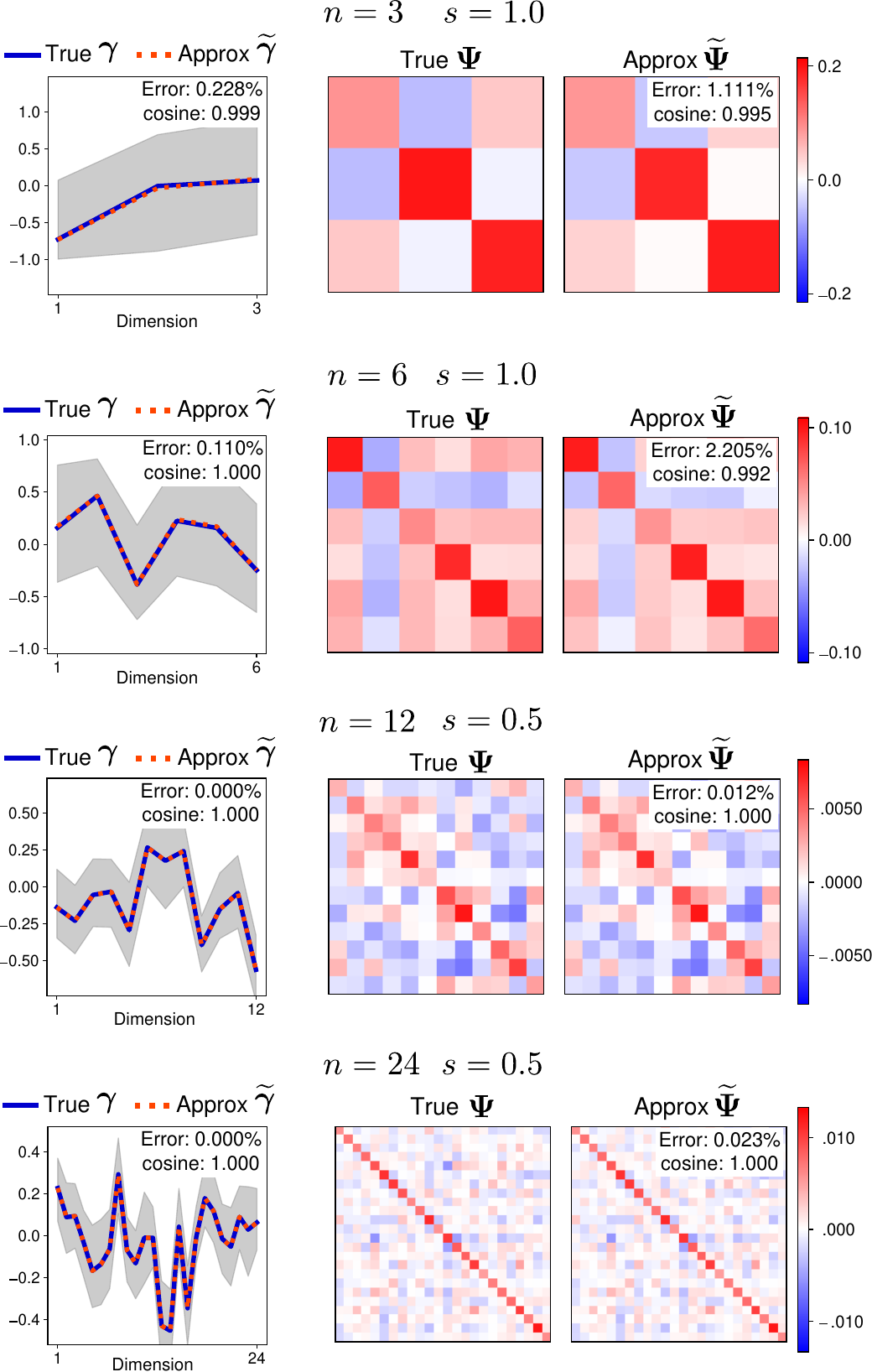}
    \caption{\textbf{Moments approximation examples for
      $\mathcal{PN}(\mub, \Sigmab)$.}
    Each row shows the true and the approximated moments for
    a projected normal distribution of a different dimensionality
    $n$ and scale $s$.
    The left column shows the true mean $\gamb$ (blue) and
    the approximated mean $\widetilde{\gamb}$ (red).
    The two means overlap almost perfectly.
    The shaded area shows the 95\% quantile interval
    of 500 samples from the distribution.
    The right column shows the true covariance $\Psib$ (left)
    and the approximated covariance $\widetilde{\Psib}$ (right).
    Insets show the approximation error and cosine similarity
    for each example.}
    \label{fig:examples1}
\end{figure}

\subsubsection{Results}

Figure~\ref{fig:examples1} shows examples of the approximated and true
moments $\gamb$ and $\Psib$ for different dimensionalities $n$ and
covariance scales $s$. The true and approximated moments are almost identical
in all cases. The largest relative squared error is smaller
than $1\%$ for all the means and $3\%$ for the
covariances. The cosine similarities are all larger than
$0.99$. These results indicate that the approximated moments
are very similar to the true moments. Visual inspection shows that
the approximated parameters are almost identical to the
true parameters, even for the largest errors.

\begin{figure}[tb]
    \centering
    \includegraphics[width=1\linewidth]{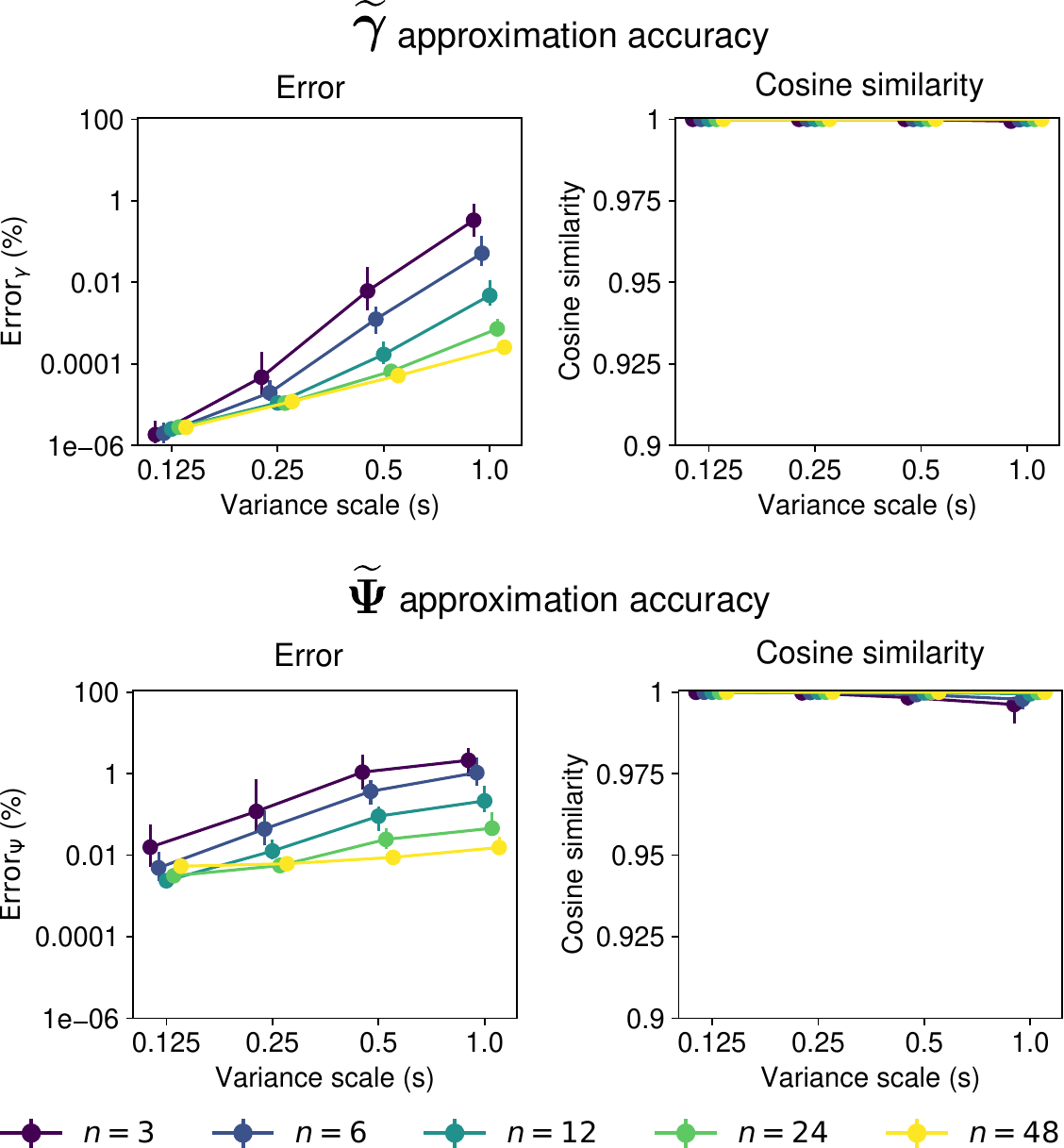}
    \caption{\textbf{Accuracy of moments approximation for
    $\mathcal{PN}(\mub, \Sigmab)$.}
    First row shows the approximation error for the
    first moment $\widetilde{\gamb}$. Second row shows the approximation
    error for the covariance $\widetilde{\Psib}$. The left panel in
    each row shows the relative mean squared error with
    a log-scale y-axis, and the right panel shows the cosine similarity.
    Points show the median and lines show the interquartile
    range for 100 samples of the parameters.
    Colors indicate the dimension $n$ of the distribution.
    The x-axis shows the scale factor $s$ of the parameter $\Sigmab$.
    }
    \label{fig:summary1}
\end{figure}

Figure \ref{fig:summary1} summarizes the approximation
errors for different dimensionalities and covariance scales.
Two features of the approximation error emerge from these
plots. First, the approximation error for both $\widetilde{\gamb}$ and
$\widetilde{\Psib}$ increases with the variance scale $s$.
Second, the error tends to decrease
with the dimensionality of the distribution.

The approximation error is small for
both $\widetilde{\gamb}$ and $\widetilde{\Psib}$, particularly in the higher
dimensional cases. For most dimensionalities and covariance scales, the median
error for both $\widetilde{\gamb}$ and $\widetilde{\Psib}$ is smaller than $1\%$,
and the median cosine similarity between the approximated and true
moments is larger than $0.99$.

Hence, the approximation method is
accurate for obtaining the first and second
moments of the projected normal distribution. The
accuracy is particularly high for larger dimensions,
and for cases where the covariance scale is not large.

\subsection{Moment matching for $\mathcal{PN}(\mub, \Sigmab)$}
\label{sec:moment_matching}

In this section, we use the moment approximations
to fit the parameters $\mub$ and $\Sigmab$ of the
projected normal distribution $\mathcal{PN}(\mub, \Sigmab)$
via moment matching.

\subsubsection{Methods}

We generated 50 samples samples of $\mub$ and $\Sigmab$
using the same procedure as in Section~\ref{sec:moment_methods}.
For a given pair of parameters $\mub$ and $\Sigmab$,
we sampled $10^7$ points from $\mathcal{PN}(\mub, \Sigmab)$,
and computed the moments $\gamb$ and $\Psib$,
which are denoted as the observed moments in the following.
The goal of the moment matching procedure is to find
the model parameters $\widehat{\mub}$ and $\widehat{\Sigmab}$
such that the moments of the corresponding distribution,
$\widetilde{\gamb}$ and $\widetilde{\Psib}$,
obtained with our approximation formulas,
are closest to the observed $\gamb$ and $\Psib$. 

To optimize the parameters $\widehat{\mub}$ and $\widehat{\Sigmab}$,
we use the mean squared difference between the observed
and the approximaed moments as a loss function:
\begin{equation}
\label{eq:optim}
  \argmin_{\widehat{\mub} \in \mathbb{S}^{n-1}, \widehat{\Sigmab} \in \mathrm{SPD}}
    (1 - \lambda) \|\widetilde{\gamb} - \gamb\|^2 + \lambda \| \widetilde{\Psib} - \Psib \|_F^2
\end{equation}
where $\lambda$ is hyperparameter that controls the relative
weight of the mean and the covariance in the loss function (see
below for details on the parameter constraints).
Unless otherwise specified, we set $\lambda=0.9$.

To evaluate the fit quality, we used the same metrics as
in Section~\ref{sec:moment_methods}. We used the
relative mean squared error between the true and the
fitted parameters:
\begin{equation*}
  \mathrm{Error}_{\mub}\% = 100 \frac{\|\widehat{\mub} - \mub\|^2}
{\|\mub\|^2}
\end{equation*}
\begin{equation*}
  \mathrm{Error}_{\Sigmab}\% = 100 \frac{\|\widehat{\Sigmab} - \Sigmab\|_F^2}{\|\Sigmab\|_F^2}
\end{equation*}
and we also used the cosine similarity:
\begin{equation*}
  \mathrm{cosine}_{\mub} = \frac{\widehat{\mub} \cdot \mub}{\|\widehat{\mub}\| \|\mub\| }
\end{equation*}
\begin{equation*}
  \mathrm{cosine}_{\Sigmab} = \frac{\mathrm{tr}\left(\widehat{\Sigmab} \Sigmab\right)}{\|\widehat{\Sigmab}\|_F \|\Sigmab\|_F}
\end{equation*}

\subsubsection{Optimization}
\label{sec:optim}

The NAdam optimizer in PyTorch \citep{paszke_automatic_2017}
was used to optimize Equation~\eqref{eq:optim}.
A cyclic schedule was used for the learning
rate. The first cycle had $80$ iterations with an initial
learning rate of $0.4$, which was multiplied by a decay
rate of $0.85$ every $5$ iterations. At the end of the $80$
iterations, the next cycle started, with a learning rate
of $0.85$ times the starting learning rate of the previous cycle.
Other than the starting learning rate, the new cycle is
identical to the first. A full optimization schedule used
$12$ such cycles.

This optimization problem is not convex, so
it is important to use a good initialization
of the parameters.
The model parameter $\widehat{\mub}$ was initialized to
$\gamb/\|\gamb\|$, and the parameter $\widehat{\Sigmab}$ was
initialized to $\Psib$.

\subsubsection{Parameterization}

The parameters $\mub$ and $\Sigmab$ of the projected normal
distribution are not uniquely determined. Scaling $\x$ by
a positive constant (with the corresponding changes in $\mub$ and
$\Sigmab$) does not change the distribution of $\y$.
Constraints are needed to make the parameters identifiable.

Previous work has placed the identifiability constraints
on the covariance matrix $\widehat{\Sigmab}$
\citep{wang_directional_2013,hernandez-stumpfhauser_general_2017,paine_elliptically_2018}.
Here, we constrain $\widehat{\mub}$ to be on the unit sphere, i.e. $||\mub||=1$,
and constrain $\widehat{\Sigmab}$ to be in the manifold of symmetric positive
definite (SPD) matrices (see Equation~\ref{eq:optim}).
To constrain $\widehat{\Sigmab}$, we use the Python package
geotorch for optimization on manifolds using trivializations
\citep{lezcano-casado_cheap_2019,lezcano_casado_trivializations_2019}.

\begin{figure}[htbp]
    \centering
    \includegraphics[width=1\linewidth]{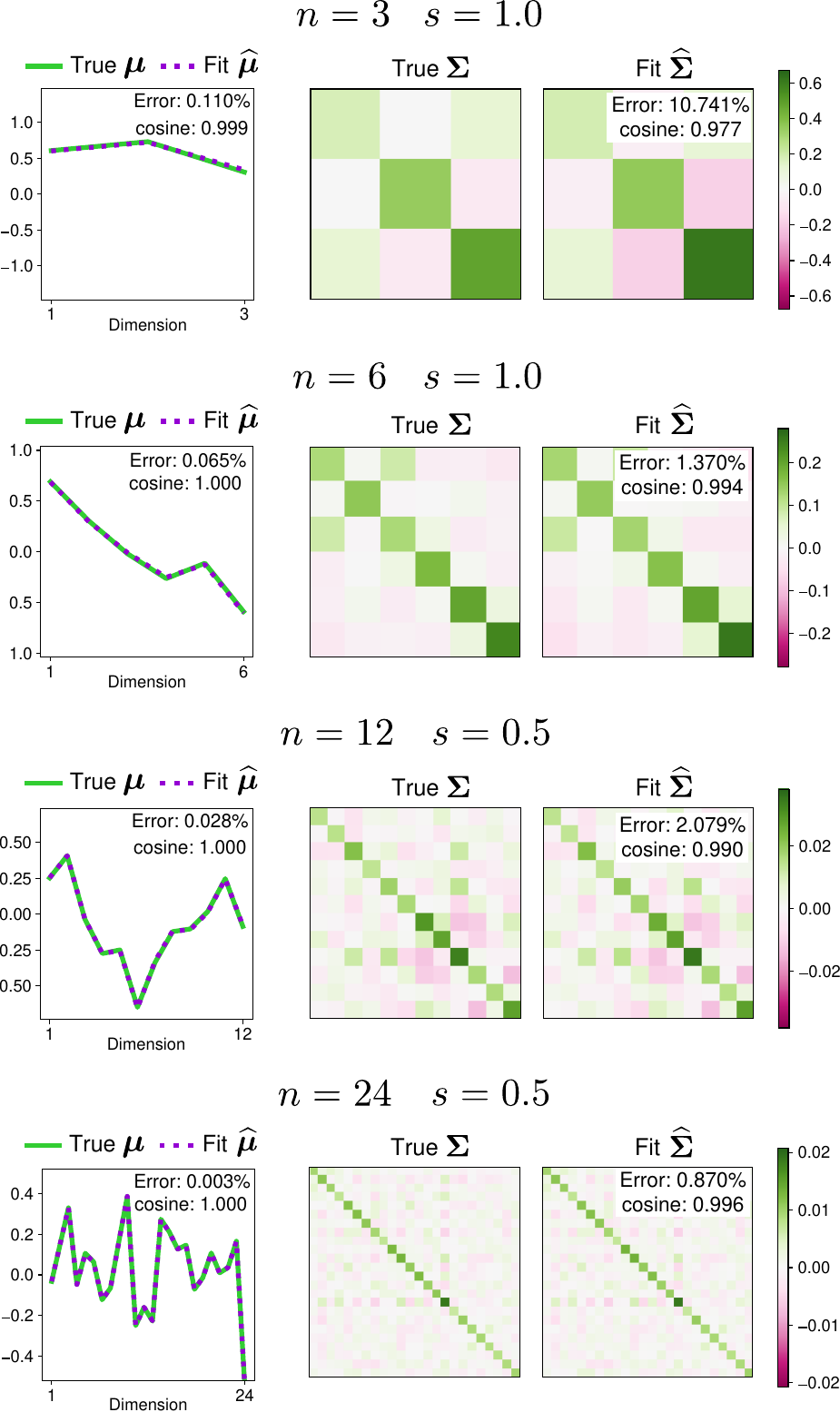}
    \caption{\textbf{Moment matching examples for $\mathcal{PN}(\mub, \Sigmab)$.}
    Each row shows the true and the fitted parameters for a
    projected normal distribution of a different dimensionality
    $n$ and scale $s$.
    The left column shows the true mean parameter $\mub$ (purple) and
    the fitted mean parameter $\widehat{\mub}$ (green). The two means
    overlap almost perfectly. The right column shows the true covariance
    parameter $\Sigmab$ (left) and the fitted covariance
    parameter $\widehat{\Sigmab}$ (right) using a color scale.
    Insets show the fit error and cosine similarity
    for each example.}
    \label{fig:examples2}
\end{figure}

\subsubsection{Results}

Figure \ref{fig:examples2} shows examples of the parameter
fits for different dimensionalities and covariance scales.
The fitted parameter $\widehat{\mub}$ is almost
identical to the true parameter $\mub$ in all examples, which is
reflected in low errors and high cosine similarities.
The fitted parameter $\widehat{\Sigmab}$ is very similar to the true
$\Sigmab$ in all examples.
The errors for $\widehat{\Sigmab}$ are larger, with an example
showing a relative errors around $10\%$. The cosine similarities
are close to $1.0$ in all cases, in line with visual inspection
showing that $\widehat{\Sigmab}$ largely captures the structure of
the true $\Sigmab$.

\begin{figure}[thbp]
    \centering
    \includegraphics[width=1.0\linewidth]{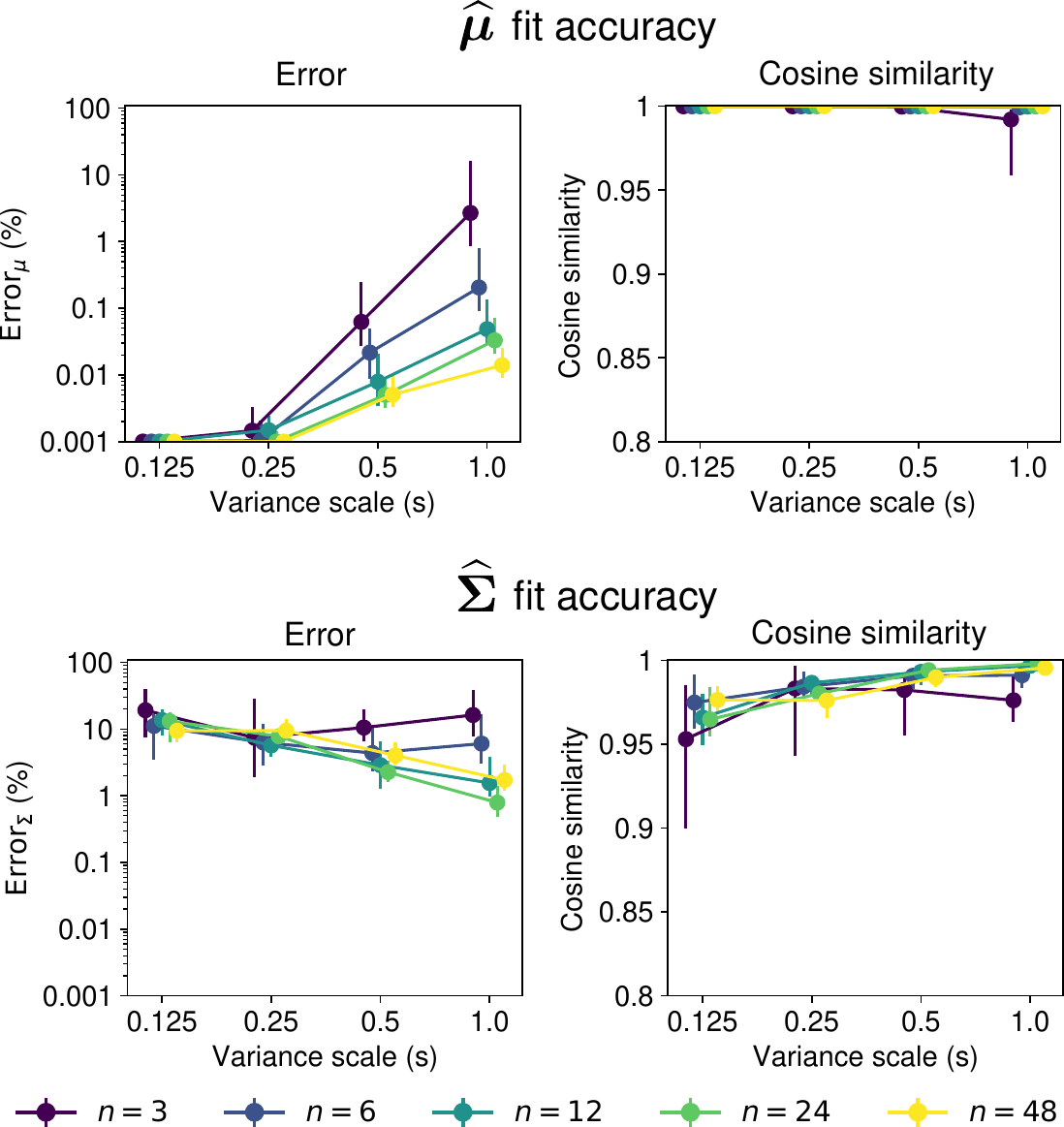}
    \caption{\textbf{Accuracy of moment matching
    for $\mathcal{PN}(\mub, \Sigmab)$.}
    First row shows the fit error for the
    mean parameter $\widehat{\mub}$. Second row shows the fit
    error for the covariance parameter $\widehat{\Sigmab}$.
    The left panel in each row shows the relative mean squared error with
    a log-scale y-axis, and the right panel shows the cosine similarity
    Points show the median and lines show the interquartile
    range for 50 samples of the parameters.
    Colors indicate the dimension $n$ of the distribution.
    The x-axis shows the scale factor $s$ of the parameter $\Sigmab$.}
    \label{fig:summary2}
\end{figure}

The median error and cosine similarity for the moment matching
fit are shown in Figure \ref{fig:summary2}.
Similar to the moment approximation
(Figure \ref{fig:summary1}), the quality of the fit tends to
increase with the dimensionality $n$.
For the mean parameter $\widehat{\mub}$, the error also increases with the
covariance scaling factor $s$. Conversely, the
error for $\widehat{\Sigmab}$ decreases as $s$ increases. The
error in the parameter fits is in general smaller
than $10\%$ for most values of $n$ and $s$, and in
many cases are in the order of $1\%$.
The cosine similarity for $\widehat{\mub}$ is close to $1.0$ for all 
values of $n$ and $s$. For $\widehat{\Sigmab}$, the median cosine
similarity is larger than $0.95$ for all values of $n$ and $s$.
Overall, the moment matching procedure is able to
estimate the model parameters with an accuracy that
depends on the dimensionality $n$ and the covariance
scale $s$, and that is high for most conditions.


\section{Generalizations of the projected normal distribution}
\label{sec:generalizations}

In this section, we present three generalizations of the
projected normal distribution. Each of these generalizations
uses a different denominator for divisive normalization,
and has a different geometrical interpretation (Figure~\ref{fig:cartoon}).
For each of the generalizations, we derive the moments
approximation, and when possible, the probability density function.

\subsection{Denominator with an additive constant: Projecting inside the unit sphere}
\label{sec:denom_c}

The first extension that we consider is the random variable
$\y = \x/\sqrt{\x^T \x + c}$, which includes a positive constant
$c$ in the denominator.
Such additive constants are a common feature of
divisive normalization models in neuroscience
\citep{albrecht_motion_1991,schwartz_natural_2001,carandini_normalization_2012}. 
We denote the distribution of this random variable as
$\mathcal{PN}_c(\mub, \Sigmab, c)$.

One key difference between $\mathcal{PN}(\mub, \Sigmab)$ and
$\mathcal{PN}_c(\mub, \Sigmab, c)$ is that the former is
defined on the surface of the unit sphere
(i.e.\ for $y$ such that $||y||=1$),
whereas the latter is defined in its interior
(i.e.\ for $y$ such that $||y||<1$) (Figure~\ref{fig:cartoon}).

\subsubsection{Probability density function}

Although the focus of this work is on the moments
of the projected normal distribution and
its extensions, we provide formulas for the
probability density function of the novel distributions
for completeness. Here, we derive the probability density
function of $\mathcal{PN}_c(\mub, \Sigmab, c)$.

Let $h(\x) = \x/\sqrt{\x^T \x + c}$ denote
the invertible transformation that maps $\x$ into $\y$.
The variables $\x$ and $\y$ have the same direction but
different norms, because $\y$ is obtained by dividing $\x$
by a positive scalar. Thus, the inverse of
$h$ is given by $h^{-1}(\y) = \y (\|\x\|/\|\y\|) = \x$.
To compute $h^{-1}(\y)$, we need to solve for
$\|\x\|$ in terms of $\y$.

Noting that $\|\y\| = \|\x\|/\sqrt{\|\x\|^2 + c}$,
solving for $\|\x\|$ results in
\begin{equation*}
  \|\x\| = \|\y\|\sqrt{\frac{c}{1 - \|\y\|^2}}
\end{equation*}
Thus, we can write the inverse transformation as
\begin{equation}
  \x = h^{-1}(\y) = \y \sqrt{\frac{c}{1 - \|\y\|^2}}
\end{equation}
The Jacobian matrix for $h^{-1}(\y)$ can be easily
computed by applying the chain rule, resulting in
\begin{equation*}
  J = \sqrt{\frac{c}{1 - \|\y\|^2}}
  \left( \mathbf{I} + \frac{\y \y^T}{1 - \|\y\|^2} \right)
\end{equation*}
A simple formula for $\det(J)$ can be obtained by using the
properties $\det(k\mathbf{A}) = k^n \det(\mathbf{A})$ for a
scalar $k$, and
$\det(\mathbf{I} + \mathbf{v}\mathbf{v}^T) = 1 + \|\mathbf{v}\|^2$,
resulting in
\begin{equation}
  \det(J) = \left(\frac{c}{1 - \|\y\|^2}\right)^{n/2}
  \left(1 + \frac{\|\y^2\|}{1 - \|\y\|^2}\right)
\end{equation}
Finally, using the change of variable formula, we obtain
\begin{align}
  \label{eq:pdf_c}
    p(\y;\mub,& \Sigmab, c) = 
    \frac{\det(J)}{\det(2\pi\Sigmab)^{\frac{1}{2}}} \nonumber \\
    & \times e^{-\frac{1}{2} \left(h^{-1}(\y) - \mub \right)^T \Sigmab^{-1}
    \left( h^{-1}(\y) - \mub\right)}
\end{align}

Like for $\mathcal{PN}(\mub, \Sigmab)$,
the formulas for the density function can be used to
fit the distribution parameters using maximum likelihood
estimation through iterative methods. We leave this
task for future work, and focus on the moments of
the distributions and moment matching procedures.

\subsubsection{Approximation to the moments of $\mathcal{PN}_c(\mub, \Sigmab, c)$}
\label{sec:moments_c}

To approximate the first and second moments of
$\y \sim \mathcal{PN}_c(\mub, \Sigmab, c)$ we use the same
Taylor expansion approach that we describe in Section~\ref{sec:moments}.
The same formulas require minor modifications
to account for the constant $c$ in the denominator.

To approximate $\gamb_i$, we consider the same
function as in Section~\ref{sec:moments_first},
$\y_i = f(\x_i,\z_i) = \x_i / \sqrt{\x_i^2 + \z_i}$
where $\z_i = \x^T \x - \x_i^2 + c$. 
The second derivatives of $f$ and thus
Equation~\eqref{eq:taylor2} are unchanged.
The expression forthe expected value of $\z_i$ is the
only change, which is now given by
\begin{equation}
\label{eq:z_c}
  \bar{\z}_i = \mathrm{tr}(\Sigmab^{i}) + \mub^{iT} \mub^i + c
\end{equation}
Therefore, to approximate $\gamb_i$, we use all the same
formulas as in Section~\ref{sec:moments_first},
except that we use Equation~\ref{eq:z_c} to obtain $\bar{\z}$.

Similarly, to approximate the second moment matrix
we consider the same function as in Section~\ref{sec:moments_second},
$\y_i \y_j = g(n_{ij},d) = n_{ij}/d$, where
$d = \x^T \x + c$ and $n_{ij} = \x_i \x_j$.
Again, the second derivatives of $g$ and thus
Equation~\eqref{eq:taylor2nd} are unchanged.
Only the expected value
of $d$ changes, which is now given by
\begin{equation}
  \label{eq:d_c}
\bar{d} = \mathrm{tr}(\Sigmab) + \mub^T \mub + c
\end{equation}
Thus, the second moment can be approximated using the
same formulas as in Section~\ref{sec:moments_second}, except
that we use Equation~\ref{eq:d_c} to obtain $\bar{d}$.


\subsection{Denominator with $\x^T \B \x$ quadratic
form: Projecting to an ellipsoid surface}
\label{sec:ellipsoid}

Next, we consider the random variable
$\y = \x/\sqrt{\x^T \B \x}$ where $\B$ is a SPD matrix
and $\x^T \B \x$ is a quadratic form of $\x$.
In neuroscience, models with such denominators have been successful
in describing complex interactions between neurons
\citep{coen-cagli_impact_2013,coen-cagli_cortical_2012}.
We denote this distribution as $\mathcal{PN}_B(\mub, \Sigmab, \B)$.

Geometrically, the variable $\y$ defined above satisfies
$\y^T \B \y = 1$. Thus $\y$ is constrained to
be on the surface of an ellipsoid defined by $\B$
(Figure~\ref{fig:cartoon}).

Importantly, the random variable
$\y \sim \mathcal{PN}_B(\mub, \Sigmab, \B)$ can be
obtained by linear transformation of a variable
$\y' \sim \mathcal{PN}(\mub', \Sigmab')$.
Let $\x' = \B^{1/2} \x$, where $\B^{1/2}$ is the matrix square
root of $\B$\footnote{The same approach can be used for
other SPD matrix decompositions, such as the Cholesky decomposition
where $\B = \L \L^T$ for a lower triangular matrix $\L$ with
positive diagonal elements.}. Then we have
\begin{align}
  \label{eq:ellipsoid}
  \y & = \frac{\x}{\sqrt{\x^T \B \x}} \nonumber \\
  &= \B^{-1/2}\frac{\x'}{\sqrt{\x'^T \x'}} \nonumber \\
  &= \B^{-1/2} \y'
\end{align}
where $\y' = \x'/\sqrt{\x'^T \x'}$.
The variable $\x'$ follows a normal distribution
$\x' \sim \mathcal{N}(\mub', \Sigmab')$,
where $\mub' = \B^{1/2} \mub$ and $\Sigmab' = \B^{1/2} \Sigmab \B^{1/2}$.
Therefore, the variable $\y'$--which is is linearly related
to $\y$--follows a projected normal distribution.
This fact can be used to obtain the moments of $\y$.

\subsubsection{Probability density function of $\mathcal{PN}_B(\mub, \Sigmab, \B)$}
\label{sec:pdf_q}

The simple linear relation between the variable
$\y \sim \mathcal{PN}_B(\mub, \Sigmab, \B)$
and the variable 
$\y' \sim \mathcal{PN}(\mub', \Sigmab')$
with a known density function
(Equation~\ref{eq:pdf1}) might suggest to the
reader that the density function of $\y$ can be
easily obtained with the change of variable formula.
However, this is not the case. Because $\y$ and $\y'$
are defined on $n-1$ dimensional surfaces,
applying the change of variable formula requires
computing how the linear transformation $\B^{-1/2}$
changes the local area element at any arbitrary point
of the sphere on which $\y'$ is defined.
This is a non-trivial task that we leave
for future work.

\subsubsection{Approximation to the moments of $\mathcal{PN}_B(\mub, \Sigmab, \B)$}
\label{sec:moments_q}

To approximate the moments of $\y$ we use the relation
$\y = \B^{-1/2} \y'$, where $\y'$ follows a
projected normal distribution.
First, the parameters $\mub'$ and $\Sigmab'$ are computed
as mentioned in Section~\ref{sec:ellipsoid}.
Then, the first and second moments of $\y'$, 
$\gamb'$ and $\Psib'$ respectively, are
computed using the formulas derived in Section \ref{sec:moments}.
Finally, the moment approximations of $\y$ are computed using
the formulas
$\gamb = \B^{-1/2} \gamb'$ and $\Psib = \B^{-1/2} \Psib' \B^{-1/2}$.


\subsection{Denominator with a quadratic form and an additive constant:
Projecting inside an ellipsoid}

Finally, we consider the random variable
$\y = \x/\sqrt{\x^T \B \x + c}$, where $\B$ is a SPD matrix
and $c$ is a positive constant.
We denote the distribution of this random variable as
$\mathcal{PN}_{Bc}(\mub, \Sigmab, \B, c)$.
Geometrically, the projected variable $\y$ defined above satisfies
$\y^T \B \y < 1$, which means that $\y$ is in
the interior of an ellipsoid defined by $\B$ (Figure~\ref{fig:cartoon}).

The variable $\y$
is related to a variable $\y' \sim \mathcal{PN}_{c}(\mub', \Sigmab', c)$
(Section~\ref{sec:denom_c}) via a linear transformation.
Defining $\x' = \B^{1/2} \x$, we have
\begin{align}
  \label{eq:ellipsoid_c}
  \y & = \frac{\x}{\sqrt{\x^T \B \x + c}} \nonumber \\
  & = \B^{-1/2}\frac{\x'}{\sqrt{\x'^T \x' + c}} \nonumber \\
  & = \B^{-1/2} \y'
\end{align}
where $\y' = \x'/\sqrt{\x'^T \x' + c}$.
The variable $\x'$ has a distribution
$\x' \sim \mathcal{N}(\mub', \Sigmab')$,
with $\mub' = \B^{1/2} \mub$ and $\Sigmab' = \B^{1/2} \Sigmab \B^{1/2}$,
implying that $\y' \sim \mathcal{PN}_{c}(\mub', \Sigmab', c)$.

\subsubsection{Probability density function}
\label{sec:pdf_qc}

The density function of $\y \sim \mathcal{PN}_{Bc}(\mub, \Sigmab, \B, c)$
can be obtained from the density function of
$\y' \sim \mathcal{PN}_{c}(\mub', \Sigmab', c)$ presented in
Equation~\ref{eq:pdf_c}. Using the change of variable formula
for linear transformations, we have
\begin{align}
  p(\y;\mub,\Sigmab,& \B,c) = \nonumber \\
  &p(\B^{-1/2}\y; \mub', \Sigmab', c) \det(\B^{1/2})
\end{align}

\subsubsection{Approximation to the moments of $\mathcal{PN}_{Bc}(\mub, \Sigmab, \B, c)$}

To approximate the moments
of $\y \sim \mathcal{PN}_{Bc}(\mub, \Sigmab, \B, c)$ we use the
fact that $\y$ is a linear transformation of
$\y'\sim \mathcal{PN}_{c}(\mub', \Sigmab', c)$
(see Section~\ref{sec:moments_q}). 
The moments can be computed using the formulas
in Section~\ref{sec:moments_c}.
First, we compute the parameters $\mub'$ and $\Sigmab'$
as described above.
Then, we compute the first and second moments of $\y'$,
$\gamb'$ and $\Psi'$, using the formulas in Section~\ref{sec:moments_c}.
Finally, we obtain the moments of $\y$ 
using the transformations
$\gamb = \B^{-1/2} \gamb'$ and
$\Psi = \B^{-1/2} \Psi' \B^{-1/2}$.

\section{Testing the moments approximation for generalizations
of the projected normal}

In this section, we test the accuracy of the moments
approximation and moment matching for the distribution
$\mathcal{PN}_{Bc}(\mub, \Sigmab, \B, c)$.
The approximation accuracy and
the moment matching accuracy for the distributions
$\mathcal{PN}_c(\mub, \Sigmab, c)$ and $\mathcal{PN}_B(\mub, \Sigmab, \B)$
are available in the Supplementary materials.
In all cases, the conclusions are the same: the
Taylor-expansion-based approximations for moments are accurate.

\subsection{Approximation accuracy for $\mathcal{PN}_{Bc}(\mub, \Sigmab, \B, c)$}

\subsubsection{Methods}

We generated 100 random samples of the parameters
$\mub$, $\Sigmab$, $\B$ and $c$
to test the accuracy of the moment approximations.
The parameters $\mub$ and $\Sigmab$ were sampled in the same way
as in Section \ref{sec:moment_methods}.

Importantly, the effect of the parameter $c$ 
depends on its scale relative to $\|\x\|^2$. At the same
time, $\mathbb{E}(\|\x\|^2)$ depends on the parameters
$n$, $\mub$, and $\Sigmab$. To make the effect of $c$
comparable across conditions,
we sampled a multiplier value $c_{mult}$ from the
exponential distribution $\mathrm{Exp}(1)$, and
then we defined the constant in the denominator
as $c=c_{mult} \mathbb{E}(\|\x\|^2)$.

Samples of $\B$ were generated with the same procedure
that was used to generate samples of $\Sigmab$ (i.e.\ we sampled
the eigenvalues independently from $\mathrm{Exp}(1)$
and the eigenvectors as described in Section~\ref{sec:moment_methods}).
Unlike $\Sigmab$, the matrix $\B$ was not scaled by
a scaling factor (i.e.\ $s$).

We used the same methods as in Section~\ref{sec:moment_methods}
to evaluate the accuracy of the moments approximation.

\subsubsection{Results}

\begin{figure}[htb]
    \centering
    \includegraphics[width=1.0\linewidth]{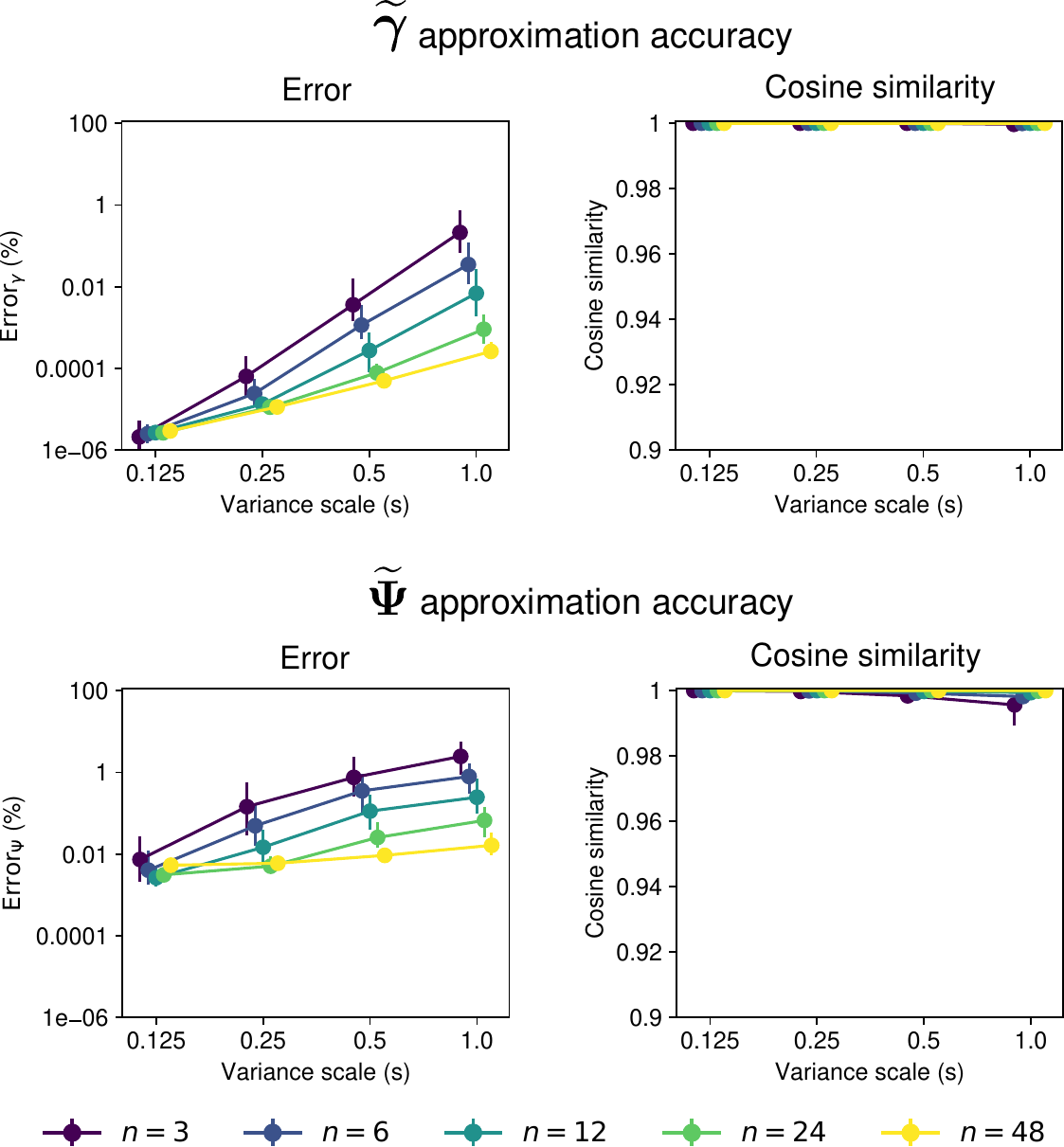}
    \caption{\textbf{Accuracy of moments approximation for
    $\mathcal{PN}_{Bc}(\mub, \Sigmab, \B, c)$.}
    First row shows the approximation error for the
    first moment $\gamb$. Second row shows the approximation
    error for the covariance $\Psib$. The left panel in
    each row shows the relative mean squared error with
    a log-scale y-axis, and the right panel shows the cosine similarity.
    Points show the median and lines show the interquartile
    range for 100 samples of the parameters.
    Points are colored by the dimension $n$ of the distribution.
    The x-axis shows the scale factor $s$ of the parameter $\Sigmab$.}
    \label{fig:summaryBc}
\end{figure}

Figure \ref{fig:summaryBc} shows the approximation
accuracy for the moments of $\mathcal{PN}_{Bc}(\mub,\Sigmab,\B,c)$.
The approximations have high accuracy, with median relative
errors for $\gamb$ and $\Psib$ smaller than $1\%$ for most values
of $n$ and $s$, and median cosine similarities larger than $0.99$.
Like for the projected normal distribution, error decreases with $n$ and
increases with the scale $s$ of the covariance
(compare to Figure~\ref{fig:summary1}).

\subsection{Moment matching for $\mathcal{PN}_{Bc}(\mub, \Sigmab, \B, c)$}

\subsubsection{Methods}

In preliminary analyses, we observed that fitting the full
parameter set $\mub$, $\Sigmab$, $\B$, and $c$ of the
distribution $\mathcal{PN}_{Bc}(\mub, \Sigmab, \B, c)$
via moment matching is difficult. This is likely
due to the large number of parameters to be estimated,
and the non-linear constraints imposed on these
parameters. Therefore, we simplified the moment matching
problem by adding further constraints to the parameters
in both the sampling and optimization steps.

First, since the focus of this section is on fitting
the denominator matrix $\B$, to reduce the total number
of free parameters we constrained the fitted
model parameter $\widehat{\Sigmab}$ to be of the form
$\widehat{\Sigmab} = \widehat{\sigma}^2 \mathbf{I}$, where
$\widehat{\sigma}^2$ is a free
parameter and $\mathbf{I}$ is the identity matrix.
The true $\Sigmab$ was sampled by
first sampling the parameter $\sigma^2$
from the uniform distribution
$\mathcal{U}(0.05 , 1.0)$, and then scaling it by the factor $s^2/n$.
Then the true covariance matrix was obtained as
$\Sigmab = \sigma^2 \mathbf{I}$.
Note that because of this constraint on $\Sigmab$, the
cosine similarity between the true and fitted covariance matrices,
$\mathrm{cosine}_{\Sigmab}$, is always $1.0$.

Second, we constrained the fitted model parameter
$\widehat{\B}$ to be of the
form $\widehat{\B} = \mathbf{I} + \widehat{b} \widehat{\vb} \widehat{\vb}$,
where $\widehat{b}$ and $\widehat{\vb}$ are free parameters.
The parameter $\widehat{b}$ is constrained to be a positive number
and $\widehat{\vb}$ is constrained to have unit norm.
The true $\B$ was obtained by sampling
$b$ as $b = 2 + \mathrm{Exp}(4)$
and $\vb$ was sampled uniformly from the unit sphere,
and then $\B$ was defined as
$\B = \mathbf{I} + b \vb \vb^T$.

Although the constraints imposed on $\widehat{\Sigmab}$ and
$\widehat{\B}$ in this section are arbitrary, for real-world
problems knowledge about $\B$ and $\Sigmab$
can be leveraged to impose useful constraints that aid
the optimization.

\subsubsection{Optimization}

Preliminary analyses showed that fit accuracy
depends on the covariance weighting
hyperparameter $\lambda$ (see Equation~\ref{eq:optim})
in way that depends on the dimensionality $n$ and
the scale $s$ of the covariance matrix.
For example, in comparing hyperparameters
$\lambda=0.9$ and $\lambda=0.98$, the former
performed much better for $n=3$ and $s=1.0$,
while the latter performed much
better for $n=48$ and $s=0.125$. Thus, for each
combination of $n$ and $s$, we used the value of
$\lambda$ that leads to the smallest median
relative squared error for the parameter $\B$.
We tested values of $\lambda=0.66$, $\lambda=0.9$,
$\lambda=0.95$ and $\lambda=0.98$.

The model parameter $\widehat{\sigma}^2$ was initialized to the value of
$\mathrm{tr}(\Psib)/n$, where $\Psib$ is the observed
covariance matrix of the data $\y$.
We initialized $\widehat{\vb}$ to the eigenvector with
the smallest eigenvalue of the observed $\Psib$, because
the eigenvector $\vb$ of $\B$ has the largest eigenvalue,
and thus the the denominator therm $\x^T \B \x$ will
attenuate the values of $\y$ in the direction
of $\vb$. $\widehat{\mu}$ was initialized to
$\widehat{\B}^{1/2} \gamma/\|\gamma\|$.

\subsubsection{Results: $\mathcal{PN}_{Bc}(\mub, \Sigmab, \B, c)$ moment matching}

\begin{figure}[htbp]
    \centering
    \includegraphics[width=0.8\linewidth]{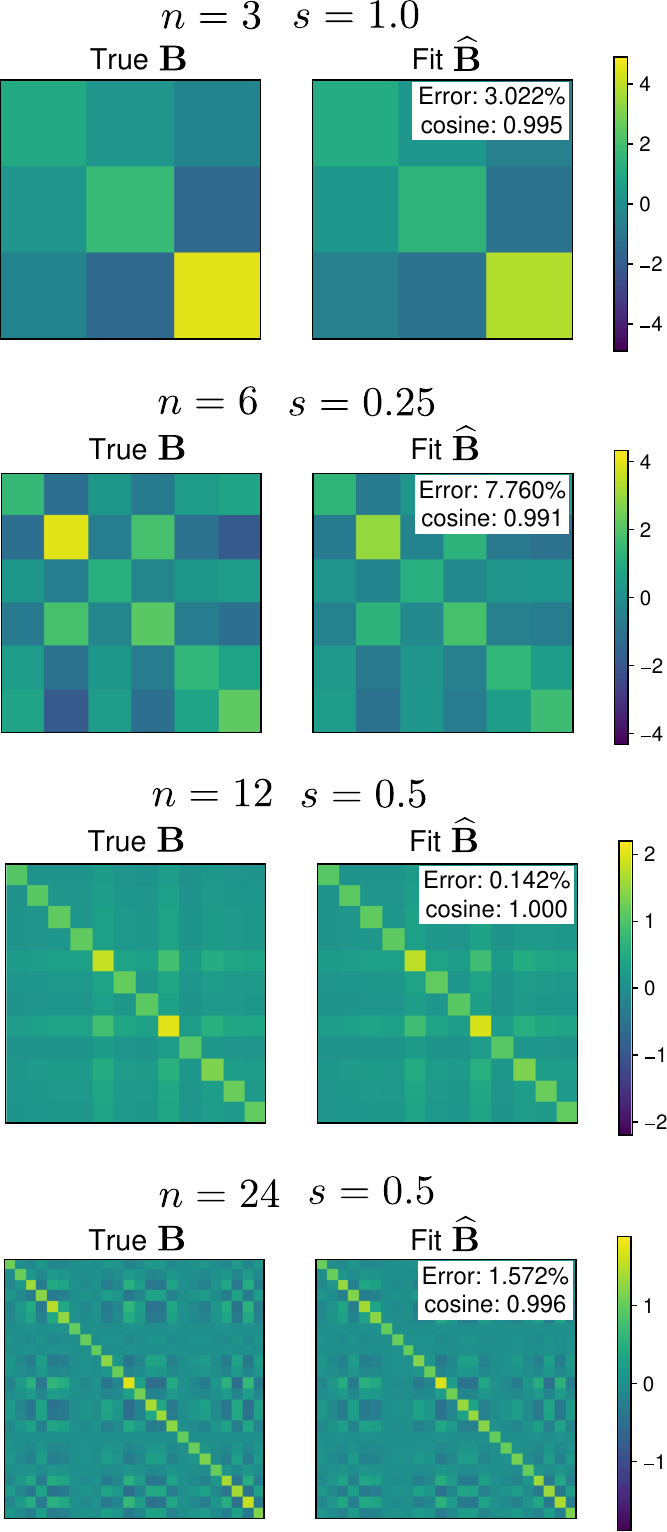}
    \caption{\textbf{Moment matching examples for the
    denominator matrix $\B$  in $\mathcal{PN}_B(\mub, \Sigmab, \B)$.}
    Each panel shows the true parameter $\B$ (left) and
    the one obtained through moment matching $\widehat{\B}$ (right)
    for different values of $n$ and $s$ (indicated above each panel).
    Insets show the error and the cosine similarity for the fitted
    matrices $\widehat{\B}$.}
    \label{fig:examples_B}
\end{figure}

Figure \ref{fig:examples_B} shows examples of the true and
the fitted $\B$ matrices for different
dimensionalities and covariance scales.
Across all examples, the cosine similarity between the true
and fitted is close to $1.0$, and the relative errors are
smaller than $10\%$, indicating that the fitted
$\widehat{\B}$ captures the structure of the true $\B$ well.

\begin{figure*}[htb]
    \centering
    \includegraphics[width=\linewidth]{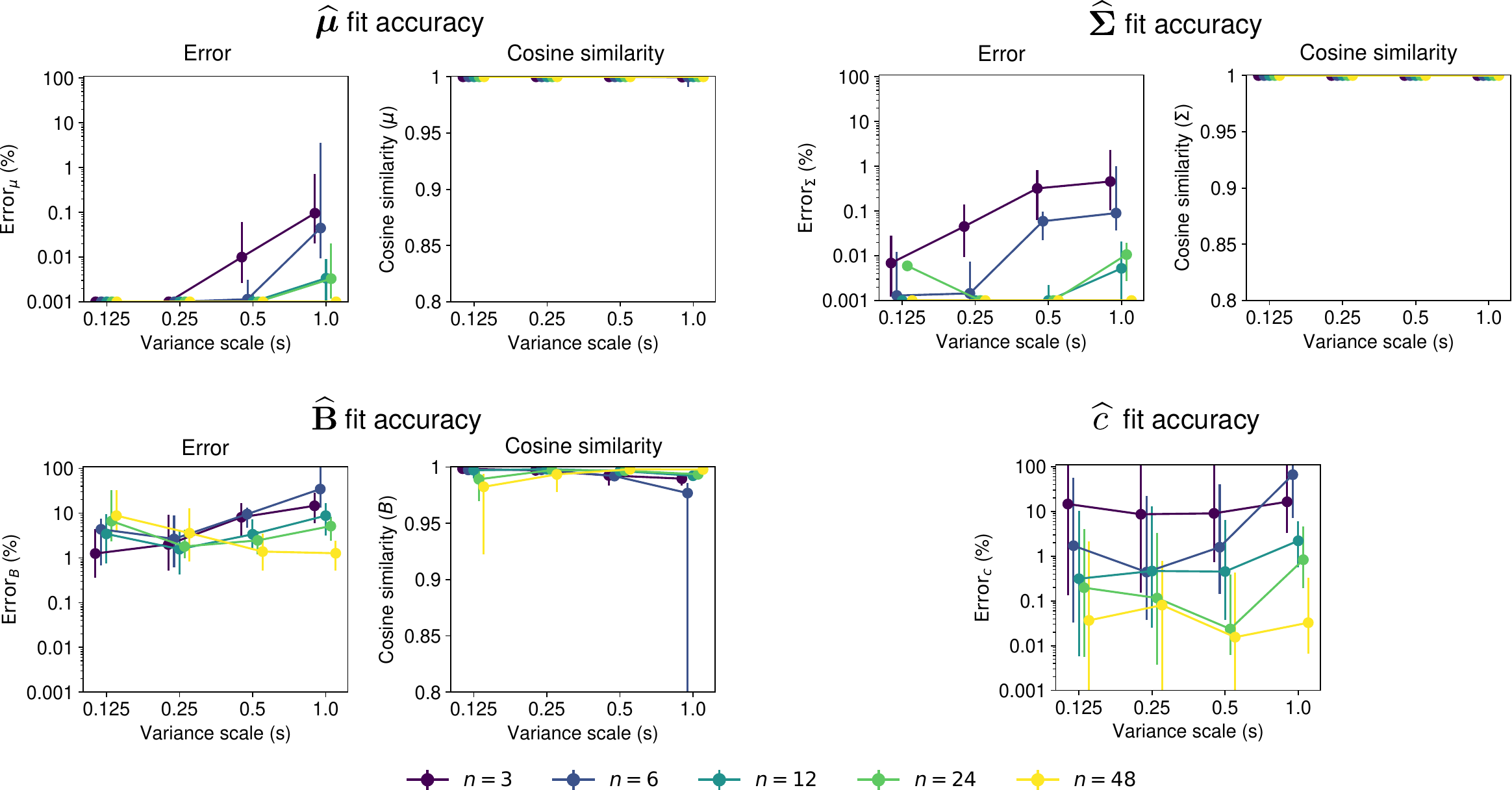}
    \caption{\textbf{Accuracy of moments matching for 
    $\mathcal{PN}_{Bc}(\mub, \Sigmab, \B, c)$.}
    First row shows the fit error for the
    mean parameter $\widehat{\mub}$ (left panel) and for the covariance
    parameter $\widehat{\Sigmab}$ (right panel).
    Second row shows the fit error for the denominator matrix $\widehat{\B}$
    (left panel) and for the denominator constant $\widehat{c}$ (right panel).
    The left plot in each panel shows the relative mean squared error with
    a log-scale y-axis, and the right panel shows the cosine similarity
    Points show the median and lines show the interquartile
    range for 50 samples of the parameters.
    Colors indicate the dimension $n$ of the distribution.
    The x-axis shows the scale factor $s$ of the parameter $\Sigmab$.}
    \label{fig:summaryBcMM}
\end{figure*}

Figure \ref{fig:summaryBcMM} shows the moment matching
fit accuracy. The accuracy for the fit is high
for for both the mean $\widehat{\mub}$ and the
covariance $\widehat{\Sigmab}$,
with median relative errors smaller than $1\%$ for all
values of $n$ and $s$.
The error for the denominator matrix $\widehat{\B}$ is also relatively
small, with median errors smaller than $10\%$ for most
values of $n$ and $s$, and median cosine similarities
larger than $0.97$ for all values of $n$ and $s$.
Interestingly, the effect of the scale $s$ is
dependent on $n$, with the error increasing with $s$
for distributions with $n=3$, and decreasing with $s$ for $n=48$.
Finally, the errors for the denominator constant $\widehat{c}$
showed a marked decrease with $n$, with median errors
in the order of $10\%$ for $n=3$, and
smaller than $1\%$ for most other conditions.

In sum, moment matching is able to accurately fit the
parameters of the distribution
$\mathcal{PN}_{Bc}(\mub, \Sigmab, \B, c)$,
including the denominator matrix $\B$ and the
denominator constant $c$ for most values of $n$ and $s$,
under the parameter constraints used in this section.


\section{Discussion}
\label{sec:discussion}

The projected normal distribution is a flexible probability
distribution on the unit sphere $\mathcal{S}^{n-1}$, that 
has seen a recent surge of interest in statistics.
In this work, we present closed-form analytic 
approximations of the first and second moments of the projected
normal distribution of arbitrary dimensionality.
These approximations have high accuracy and are computationally
efficient.

Furthermore, we show that the approximations can be used to
fit the parameters of the distribution through a moment matching
procedure, resulting in accurate parameter estimates across a
wide range of values of dimensionality $n$, and of scale
$s$ of the covariance matrix $\Sigmab$.
The moment approximation and moment matching procedures
are a useful addition to the set of statistical tools recently
developed for the projected normal distribution.
The simplicity of the moment matching procedure 
allows for fast and efficient fitting of the parameters
that should make the projected normal distribution
more accessible in practice.
We make these approximations and the moment matching procedures
available as a Python package named `projnormal'.

We also introduced three different generalizations of the
projected normal distribution motivated by an
influential model of neural activity from
computational neuroscience, the divisive normalization model
\citep{carandini_normalization_2012,goris_response_2024,herrera-esposito_analytic_2024}.
Geometrically, while the projected normal distribution
projects the random variable onto the unit sphere,
these generalizations of the projected normal
project the random variable
to the surface or to the interior of an ellipsoid. For some
of these generalizations, we derive the probability density
function of the distribution. For all of the generalizations, we
find moment approximations, and show that moment matching can be
used to fit the parameters of the distributions.

Divisive normalization models have been successful in
describing and predicting neural responses in a wide range
of systems and experimental setups in neuroscience
\citep{schwartz_natural_2001,coen-cagli_cortical_2012,
carandini_normalization_2012,goris_response_2024}.
The projected normal distribution and
its generalizations presented here should provide a natural
framework to extend divisive normalization models to
account for the joint of effects of noisy inputs
and normalization on the statistical properties of neural response.
The moments of the projected normal distribution
and its generalizations are particularly relevant given recent
progress in estimating the moments of neural population responses
across unobserved experimental conditions
\citep{nejatbakhsh_estimating_2023,ding_information_2023},
producing rich sets of moments that can be used to
fit the parameters of the distributions, for which
the underlying data is not available.

Some challenges remain in the application of the
projected normal distribution and its generalizations
to real-world problems. For the generalizations
using a quadratic form $\x^T \B \x$ in the denominator,
the matrix $\B$ adds a large number of parameters to the
distribution, rendering optimization difficult.
In this work, we used a constraint on $\B$ to reduce the
number of parameters and to improve the optimization, and
also constrained the covariance matrix $\Sigmab$ to
be isotropic. Future work should explore ways to improve
the optimization of the parameters $\B$ and $\Sigmab$.
Other methods for parameterization, initialization, and/or
optimization should be helpful towards these efforts.
For concrete applications, knowledge about the problem
can be used to add problem-specific constraints that can
aid the optimization while still allowing for a flexible
representation of the data.

\backmatter

\bmhead{Acknowledgements}

This work was supported by the National Eye Institute and the Office of Behavioral and Social Sciences Research, National Institutes of Health Grant R01-EY028571 to J.B.

\begin{appendices}

\section{Moments of quadratic forms of Gaussian random variables}
  \label{sec:quadratics}

For convenience, we present here some useful formulas for the
moments of quadratic forms of Gaussian random variables,
that can be found in \citep{mathai_quadratic_1992}
Chapter 3.2.

Let $\x \in \mathbb{R}^n$ be a Gaussian random variable
$\x \sim \mathcal{N}(\mub, \Sigmab)$, where
$\mub \in \mathbb{R}^n$ is the mean vector and
$\Sigmab \in \mathbb{R}^{n \times n}$ is the covariance matrix.
Let $\mathbf{M} \in \mathbb{R}^{n\times n}$ and
$\mathbf{K}\mathbb{R}^{n\times n}$ be symmetric positive
definite (SPD) matrices, and $\mathbf{b} \in \mathbb{R}^n$ be a vector.
Then, $q = \x^T \mathbf{M} \x$ and $p = \x^T \mathbf{K} \x$
are two quadratic forms of $\x$, and $l = \mathbf{b}^T \x$ is a
linear function of $\x$.
The following formulas are useful for computing
the moments of the projected normal distribution

\begin{align}
  \bar{q} &= \mathrm{tr}(\mathbf{M}\Sigmab) + \mub^T \mathbf{M} \mub \label{eq:sup1} \\
  \var(q) &= 2 \mathrm{tr}(\mathbf{M}\Sigmab \mathbf{M}\Sigmab) +
    4 \mub^T \mathbf{M} \Sigmab \mathbf{M} \mub \label{eq:sup2} \\
  \cov(q,p) &= 2\mathrm{tr}(\mathbf{M}\Sigmab \mathbf{K} \Sigmab) +
  4 \mub^T \mathbf{M} \Sigmab \mathbf{K} \mub \label{eq:sup3} \\
  \cov(q,l) &= 2 \mub^T \mathbf{M} \Sigmab \mathbf{b} \label{eq:sup4}
\end{align}

For the approximation of $\gamb_i$, we used the
variables $\x_i$ and $\z_i = \x^T\x - \x_i^2$.
We can consider $\x_i$ as a linear function of $\x$ given by
$\x_i = \mathbf{e}_i^T \x$, where $\mathbf{e}_i$ is the canonical
$i$ vector (i.e.\ a vector with $1$ in position $i$
and $0$ elsewhere). Note
that $\z_i$ is the sum of the squares of all the elements
of $\x$ except for the $i$'th element. We can define
$\z$ as the quadratic form $\z = \x^T \mathbf{I}_{-i} \x$,
where $\mathbf{I}_{-i}$ is the same as the identity matrix
everywhere, except in the $i$'th diagonal element
where it is $0$. Then, we can use the formulas above
to obtain the moments required to approximate $\gamb_i$.

Using Equation~\ref{eq:sup1} with $\mathbf{M} = \mathbf{I}_{-i}$,
we get the formula for $\bar{\z}_i$ in the main text.
Using Equation~\ref{eq:sup2} with $\mathbf{M} = \mathbf{I}_{-i}$,
we get the formula for $\var(\bar{\z}_i)$ in the main text.
Finally, using Equation~\ref{eq:sup4} with $\mathbf{M} = \mathbf{I}_{-i}$
and $\mathbf{b} = \mathbf{e}_i$, we get the formula for
$\cov(\x_i, \z_i)$ in the main text.

Similarly, for the approximation of the second moment
matrix element $\mathbb{E}[\y_i \y_j]$, we used
the variables $n_{ij} = \x_i \x_j$ and 
$d = \x^T \mathbf{I} \x$. We can consider $n_{ij}$ as a
quadratic form of $\x$ given by
$n_{ij} = \x^T \mathbf{A}^{ij} \x$, where
$\mathbf{A}^{ij}$ is a matrix with entries
$i,j$ and $j,i$ equal to $1/2$ in the case $i \neq j$,
and $1$ in the case $i=j$, and $0$ elsewhere.

Using Equation~\ref{eq:sup1} with $\mathbf{M} = \mathbf{A}^{ij}$,
we obtain the formula for $\bar{n}_{ij}$ in the main text.
Using Equation~\ref{eq:sup1} with $\mathbf{M} = \mathbf{I}$
we obtain the formula for $\bar{d}$ in the main text.
Using Equation~\ref{eq:sup2} with $\mathbf{M} = \mathbf{I}$,
we obtain the formula for $\var(d)$ in the main text.
Finally, using Equation~\ref{eq:sup3} with
$\mathbf{M} = \mathbf{A}^{ij}$ and $\mathbf{K} = \mathbf{I}$,
we obtain the formula for $\cov(n_{ij}, d)$ in the main text.

\section{Vectorized formulas for moments approximation}
\label{sec:vectorized}

\subsection{Vectorized formulas for $\gamb$}

In Section~\ref{sec:moments_first} of the main text, we provide
formulas to approximate the $i$'th element of the first moment
$\gamb_i$ of the projected normal distribution. To
compute the approximation to the full vector $\gamb$,
the approximation for each element $i$ can be computed
separately. However, the different elements $\gamb_i$ share
many of the same computations, and thus it is possible to
compute $\gamb$ in an efficient vectorized way.

First, we define the following vectors containing the
required moments for every $\z_i$:
\begin{align*}
  \bar{\z} &= \left[\bar{\z}_1, \ldots, \bar{\z}_n\right] \nonumber \\
  \var(\z) &= \left[\var(\z_1), \ldots, \var(\z_n)\right] \nonumber \\
  \cov(\x,\z) &= \left[\cov(\x_1,\z_1), \ldots, \cov(\x_n,\z_n)\right]
\end{align*}
A vectorized formula for $\gamb$ can be obtained as follows
(compare to Equation~\ref{eq:taylor2} in the main text):
\begin{align*}
  \gamb &\approx
    \frac{\mub}{(\mub^2 + \bar{\z})^{\frac{1}{2}}} +
    \frac{\mathrm{diag}(\Sigmab)}{2} \odot \left( \frac{-3  \mub \odot \bar{\z}}{
      ( \mub^2 + \bar{\z})^{5/2}} \right)
     \nonumber \\
    & \quad + \frac{\var(\z) }{2} \odot
    \left( \frac{3 \mub}{4( \mub^2 + \bar{\z})^{5/2}} \right) \nonumber \\
    & \quad + \cov(\x,\z) \odot \left( \frac{ \mub^2 - \frac{1}{2}\bar{\z}}{( \mub^2 + \bar{\z})^{5/2}} \right)
\end{align*}
where $\mathrm{diag}(\Sigmab)$ is a vector with the diagonal
elements of $\Sigmab$, $\odot$ is the Hadamard product
and, abusing the notation, the powers and divisions involving
vectors are element-wise.

Then, efficient vectorized formulas can be obtained for
the vectors $\bar{\z}$, $\var(\z)$, and $\cov(\x,\z)$.
For instance, consider the formula for $\bar{\z}_i$ in the main text,
\begin{equation*}
  \bar{\z}_i = \mathrm{tr}(\Sigmab^i) + \mub^{iT}\mub^{i}
\end{equation*}
where $\Sigmab^i$ is the covariance matrix $\Sigmab$
with the $i$'th diagonal element set to $0$, and
$\mub^i$ is the mean vector $\mub$ with the $i$'th element
set to $0$. This is equivalent to the formula where we
use the full covariance matrix $\Sigmab$ and
the full mean vector $\mub$, and then subtract the
terms corresponding to the $i$'th element:
\begin{equation*}
  \bar{\z}_i = \mathrm{tr}(\Sigmab) + \mub^{T}\mub - \Sigmab_{ii} - \mub_i^2
\end{equation*}
Thus, to compute each $\bar{\z}_i$, we can compute the
shared terms $\mathrm{tr}(\Sigmab) + \mub^{T}\mub$,
and then subtract the $i$-specific terms $\Sigmab_{ii} + \mub_i^2$.
This leads to the vectorized formula
\begin{equation*}
  \bar{\z} = \mathbf{1} (\mathrm{tr}(\Sigmab) + \mub^T\mub) - (\mathrm{diag}(\Sigmab) + \mub^2)
\end{equation*}
where $\mathbf{1}$ is a vector of ones of length $n$,
and the squaring of $\mub$ is element-wise.

Applying this approach to the formulas
for $\var(\z_i)$ and $\cov(\x_i, \z_i)$ in the main text
and using some algebraic manipulations, the
following vectorized formulas can be obtained:
\begin{align*}
  \var(\bar{\z}) &= \mathbf{1} \left(2 \mathrm{tr}(\Sigmab \Sigmab) +
    4 \mub^T \Sigmab \mub\right) \nonumber \\
  & \quad - 2 \left(2 \, \mathrm{diag}(\Sigmab\Sigmab) - \mathrm{diag}(\Sigmab)^2\right) \nonumber \\
  & \quad - 4 \left(2\mub \odot \Sigmab \mub - \mub^2 \odot \mathrm{diag}(\Sigmab)\right) \\
  \cov(\x, \z) & =
  2 ( \mub^T  \Sigmab - \mub \odot  \mathrm{diag} (\Sigmab))
\end{align*}

Combining these vectorized formulas for the moments
of the variables $\z_i$ with the vectorized formula for $\gamb$,
we can efficiently compute the approximation to
the first moment $\gamb$ of the projected normal distribution.

\subsection{Vectorized formula for $\mathbb{E}\left[\y \y^T\right]$ approximation}

To obtain a vectorized formula for the second moment matrix
$\mathbb{E}\left[\y \y^T\right]$ of the projected normal,
we can first define the matrices $\bar{\N} \in \mathbb{R}^{n\times n}$
and $\cov(\N,d) \in \mathbb{R}^{n \times n}$,
with elements $\bar{\N}_{ij} = \bar{n}_{ij}$ and
$\cov(\N,d)_{ij} = \cov(n_{ij},d)$. Then we can define the following
vectorized formula for the second moment matrix
(compare to Equation~\ref{eq:taylor2nd} in the main text):
\begin{equation*}
\label{eq:taylorsm}
    \mathbb{E}[ \y\y^T ] \approx \frac{\bar{\N}}{\bar{d}} \odot
    \left( \mathbf{1} - \frac{\cov(\N,d)}{\bar{\N}\bar{d}} +
    \frac{\mathbf{1} \cdot \var(d)}{\bar{d}^2}
    \right)
\end{equation*}
where the divisions between vectors are element-wise.

Then, like for the first moment $\gamb$, the moments
of the auxiliary variable $\bar{\N}$ and
$\cov(\N,d)$ can be computed using efficient
vectorized formulas. For example, consider the formula for
$\bar{n}_{ij}$ in the main text. It is clear that
\begin{equation*}
\bar{\N} = \Sigmab + \mub \mub^T
\end{equation*}

Then, consider the formula for $\mathrm{cov}(n_{ij},d)$ in the
main text. It is easy to show that the first term
$2 \mathrm{tr}(\Sigmab\mathbf{A}^{ij} \Sigmab)$ is equal to
$2 \Sigmab_{i:} \Sigmab_{:j}$, where $\Sigmab_{i:}$
is the $i$'th row of $\Sigmab$ and $\Sigmab_{:j}$ is the $j$'th column.
Thus, the matrix product $2 \Sigmab \Sigmab$ has the
corresponding element $2 \Sigmab_{i:} \Sigmab_{:j}$
at each position $(i,j)$. Applying similar algebraic
manipulations to the second term, we obtain the
following vectorized formula for $\cov(\N,d)$:
\begin{equation*}
  \cov(\N,d) = 2 \left(\Sigmab \Sigmab + \mub \mub^T \Sigmab + \Sigmab \mub \mub^T \right)
\end{equation*}

Then, using the vectorized formulas for $\bar{\N}$ and
$\cov(\N,d)$, we can compute the second moment matrix
$\mathbb{E}[\y \y^T]$ of the projected normal distribution
using the vectorized formula above.

\section{Exact moments of $\mathcal{PN}(\mub, \sigma^2 \mathbf{I})$}
\label{sec:exact_moments}

\subsection{Exact solution for $\gamb$}

For the case where $\Sigmab = \mathbf{I}\sigma^2$, an exact
closed-form formula can be derived for $\gamb$. To obtain
the closed-form formula for $\gamb$, we first note that
because of the symmetry of $\x$ around $\mub$, $\gamb$
will be parallel to $\mub$. Therefore, we only need to
solve for the expected
value of the projection of $\y$ onto the direction of $\mub$

\begin{align}
 \mathbb{E}& \left[ \y^T \cdot \frac{\mub}{\|\mub\|}\right] =
  \mathbb{E} \left[ \frac{\x^T \mub}{\|\x\| \|\mub\|}\right] = \nonumber \\
   & \frac{1}{(2\pi \sigma^2)^{\frac{n}{2}}} \int_{\mathbb{R}^n}
e^{-\frac{1}{2\sigma^2}(\x-\mub)^T(\x-\mub)} \frac{\x^T \mub}{\|\x\| \|\mub\|} d\x = \nonumber \\
    & \frac{1}{(2\pi \sigma^2)^{\frac{n}{2}}}  e^{-\frac{\mub^2}{2\sigma^2}}
\int_{\mathbb{R}^n}
e^{-\frac{r^2}{2\sigma^2}} e^{\frac{r\|\mub\|}{\sigma^2} \mathrm{cos}(\theta)}
\mathrm{cos}(\theta) d\x = \nonumber \\
    & \frac{1}{(2\pi \sigma^2)^{\frac{n}{2}}}  e^{-\frac{\mub^2}{2\sigma^2}}
\int_{0}^{\infty} dr \int_{0}^{\pi } r \mathrm{A}_{n-2}(r\sin(\theta)) \nonumber \\
  & \quad e^{-\frac{r^2}{2\sigma^2}} e^{\frac{r\|\mub\|}{\sigma^2}\cos(\theta)} 
\cos(\theta) d\theta
\end{align}
where
\begin{equation*}
  \mathrm{A}_{n-2}(r\sin(\theta)) = \frac{2\pi^{(n-1)/2}}{\Gamma\left(\frac{n-1}{2}\right)}\,r^{n-2} \sin^{n-2}(\theta)
\end{equation*}
is the area of the $n-2$ sphere of radius $r\sin(\theta)$,
and $\theta$ is the angle between $\x$ and $\mub$.
In the first step above, 
we expanded the term $(\x-\mub)^T(\x-\mub)$, we used the identity
$\frac{\x \cdot \mub}{\|\x\| \|\mub\|} = \mathrm{cos}(\theta)$,
where $\theta$ is the angle between $\x$ and $\mub$, and
denoted the norm of $\|\x\|$ as $r$, refering to the radius
from the origin to the point $\x$.
In the second step, we decomposed the integral
into polar coordinates using the angle $\theta$ between $\x$ and $\mub$.
The angular components other than $\theta$ are
encompassed by the area $\mathrm{A}_{n-2}(r\sin(\theta))$.
This is because for a fixed $r$ and $\theta$, all the points
defined by the remaining angular directions form an $n-2$ sphere
of radius $r\sin(\theta)$ where all the points
have the same probability (because of symmetry around $\mub$),
and the integrand does not depend on these directions.
Thus, the integral over the remaining angular directions is
equal to the area of the $n-2$ sphere of radius $r\sin(\theta)$.

Grouping all $\theta$-dependent terms in the integral
and substituting $\cos(\theta)=z$,
$\sin^{n-2}(\theta)=(1-z^2)^{\frac{n-2}{2}}$ and
$d\theta = -(1-z^2)^{-\frac{1}{2}} dz$, we can
evaluate the integral with respect to $\theta$ using an integral
representation of the hypergeometric function ${}_0F_1$
\begin{align*}
  \int_{1}^{-1} & - z e^{\left(\frac{r\|\mub\|}{\sigma^2}\right)z}
    (1-z^2)^{\frac{n-2}{2}-\frac{1}{2}} dz  = \nonumber \\
  & \frac{\sqrt{\pi}}{2}
 \frac{\Gamma\left( \frac{n-1}{2} \right)}{\Gamma\left( \frac{n}{2}+1 \right)} \frac{\|\mub\| r}{\sigma^2}
    {}_0F_1\left(\frac{n}{2}+1; \left( \frac{\|\mub\| r}{2\sigma^2}\right)^2 \right)
\end{align*}

Using this integral representation,
moving outside of the integral the terms that do not depend
on $r$, and simplifying, we obtain the following
\begin{align*}
  \mathbb{E} & \left[ \frac{\x^T \mub}{\|\x\| \|\mub\|}\right]  =
 \frac{1}{(2\sigma^2)^{\frac{n}{2}}}
 \frac{1}{\Gamma\left( \frac{n}{2}+1 \right)}
 \frac{\|\mub\|}{\sigma^2} e^{-\frac{\|\mub\|^2}{2\sigma^2}} \nonumber \\
  & \quad \int_{0}^{\infty} r^n e^{-\frac{r^2}{2\sigma^2}}
{}_0F_1\left(\frac{n}{2}+1; \left( \frac{\|\mub\| r}{2\sigma^2}\right)^2 \right) dr
\end{align*}

Next, substituting $h=\frac{r^2}{2\sigma^2}$ and
$dr = \frac{\sigma^2}{\sqrt{2h}}dh$, simplifying, and using
the integral representation
${}_1F_1(a;b;h)\Gamma(a) = \int_{0}^{\infty} e^{-t} t^{a-1} {}_0F_1\left(b;ht\right)dt$
we obtain

\begin{align}
  \label{eq:projection_exact}
  \mathbb{E}&\left[ \y^T \cdot \frac{\mub}{\|\mub\|}\right]  = \frac{1}{(2\sigma^2)^{\frac{1}{2}}} \frac{1}{\Gamma\left( \frac{n}{2}+1 \right)}
 \|\mub\| e^{-\frac{\|\mub\|^2}{2\sigma^2}} \nonumber \\
    & \quad \quad \times \int_{0}^{\infty} h^{\frac{n}{2}-\frac{1}{2}} e^{-h}
    {}_0F_1\left(\frac{n}{2}+1; \frac{\mub^2}{2\sigma^2}h \right) dh \nonumber \\
  & = \frac{1}{(2\sigma^2)^{\frac{1}{2}}}
\frac{\Gamma\left(\frac{n+1}{2} \right)}{\Gamma\left( \frac{n}{2}+1 \right)}
 \|\mub\|e^{-\frac{\|\mub\|^2}{2\sigma^2}} \nonumber \\
  & \quad \quad \times {}_1F_1\left(\frac{n+1}{2}; \frac{n}{2}+1; \frac{\|\mub\|^2}{2\sigma^2}\right) \nonumber \\
  & = \frac{1}{(2\sigma^2)^{\frac{1}{2}}}
\frac{\Gamma\left(\frac{n+1}{2} \right)}{\Gamma\left( \frac{n}{2}+1 \right)}
 \|\mub\| {}_1F_1\left(\frac{1}{2}; \frac{n}{2}+1; - \frac{\|\mub\|^2}{2\sigma^2}\right)
\end{align}
where in the last step we used Kummer's transformation
${}_1F_1(a;b;m) = e^{m}{}_1F_1(b-a;b;-m)$. Thus, multiplying
the result of Equation~\ref{eq:projection_exact} by $\mub/\|\mub\|$,
we obtain the closed-form formula for $\gamb$:
\begin{align}
\gamb = & \frac{1}{(2\sigma^2)^{\frac{1}{2}}}
\frac{\Gamma\left(\frac{n+1}{2} \right)}{\Gamma\left( \frac{n}{2}+1 \right)} \nonumber \\
  & \quad \times {}_1F_1\left(\frac{1}{2}; \frac{n}{2}+1; - \frac{\|\mub\|^2}{2\sigma^2}\right) \mub
\end{align}

\subsection{Exact solution for $\mathbb{E}[yy^T]$}

As mentioned in Section~\ref{sec:quadratics}, 
each element of $\mathbb{E}\left[\y\y^T\right]$
is the expectation of a ratio
of quadratic forms in random variables $\mathbb{E}[n_{ij}/d]$ 
where $n_{ij}=\x^T A^{ij} \x$ and $d=\x^T \x$, where $\x \sim \mathcal{N}(\mub, \Sigmab)$.
For the special case where $\Sigmab = \sigma^2\mathbf{I}$,
the expectation of this ratio of quadratic forms has an
exact solution \citep{smith_expectations_1993,smith_comparing_1996}
\begin{align}
\label{eq:exact_sm}
  \mathbb{E}&\left[\frac{\x^T \mathbf{M} \x}{\x^T \x} \right] =
   \frac{ \mathrm{tr}(\mathbf{M})}{n} \, {}_{1}F_1\left(1; \frac{n}{2}+1; -\frac{\|\mub\|^2}{2\sigma^2}\right) \nonumber \\
  & \quad \quad  + \frac{\mub^T \mathbf{M} \mub}{n+2} \, {}_{1}F_1\left(1; \frac{n}{2}+2; -\frac{\|\mub\|^2}{2\sigma^2}\right)
\end{align}

Thus, the value of $\mathbb{E}\left[\y_i\y_j^T\right]$ can be obtained
by substituting $\mathbf{M} = \mathbf{A}^{ij}$ in Equation~\ref{eq:exact_sm},
with $\mathbf{A}^{ij}$ being the matrix defined in
Section~\ref{sec:quadratics}. Noting that 
$\mathrm{tr}(\mathbf{A}^{ij})$ is $1$ when $i=j$ and
$0$ otherwise, and that $\mub^T \mathbf{A}^{ij} \mub = \mub_i \mub_j$,
it is easy to verify the following matrix formula
\begin{align}
  \mathbb{E}&\left[\y \y^T \right] = \frac{1}{n} \, {}_{1}F_1\left(1; \frac{n}{2}+1; -\frac{\|\mub\|^2}{2\sigma^2}\right) \mathbf{I} \nonumber \\
     & \quad + \frac{1}{n+2} \, {}_{1}F_1\left(1; \frac{n}{2}+2; -\frac{\|\mub\|^2}{2\sigma^2}\right) \mub\mub^T
\end{align}

\end{appendices}


\bibliography{sn-bibliography}

\end{document}